\documentstyle[12pt]{article}
\def\Vubabs{\vert V_{ub}\vert}

\def\Vcbabs{\vert V_{cb} \vert}

\def\Vtdabs{\vert V_{td} \vert}

\def\Vtsabs{\vert V_{ts} \vert}

\def\mt{m_t}

\def\pp{{\prime\prime}}
\def\pg{}
\newcommand{\ud}{\uparrow\!\downarrow}                %hyp. tang
\newcommand{\du}{\downarrow\!\uparrow}                %hyp. tang
\newcommand{\go}[1]{\gamma^{#1}}
\newcommand{\gu}[1]{\gamma_{#1}}

\newcommand{\ra}{\rightarrow}

\newcommand{\BBKSTAR}{BR(B \ra K^\star + \gamma)}
\newcommand{\BKSTAR}{B \ra K^\star + \gamma}
\newcommand{\BGAMAXS}{B \ra X _{s} + \gamma}
\newcommand{\BGAMAXD}{B \ra X _{d} + \gamma}
\newcommand{\BBGAMAXS}{BR(B \ra  X _{s} + \gamma)}
\newcommand{\BBGAMAXD}{BR(B \ra  X _{d} + \gamma)}

\newcommand{\BBGAMAKSTAR}{BR(B \ra  K^{\star} + \gamma)}

\newcommand{\BGAMAKSTAR}{B \ra  K^{\star} + \gamma}

\newcommand{\BGAMAS}{b \ra s + \gamma}
\newcommand{\BGAMAD}{b \ra d + \gamma}

\newcommand{\bsgam}{\ $b \to s+ \gamma$}

\newcommand{\bsggam}{\ $b \to s+ \gamma+ g$}

\newcommand{\fbb}{f^2_{B_d}B_{B_d}}
\newcommand{\fbbs}{f^2_{B_s}B_{B_s}}
\begin{document}
\thispagestyle{empty}
{}~\vspace*{-4cm}
\begin{flushright}
CERN--TH.7118/93\\
MPI--Ph/93--97\\
DESY 93--193
%December 1993
\end{flushright}
\vskip0.5cm
\begin{center}
{\Large \bf
Exclusive Radiative $\bf B$-Decays in the\\ Light-Cone QCD Sum Rule
Approach \\ }
\vskip1cm
 {\large A.~Ali}$\footnote{On leave of absence from DESY, Hamburg, FRG.}$
\vskip0.2cm
       Theory Division, CERN  \\
       CH-1211 Geneva 23, Switzerland \\
\vskip0.5cm
 {\large V.M.~Braun} $\footnote { On leave of absence from
St.Petersburg Nuclear
Physics Institute, 188350 Gatchina, Russia.}$ \\
\vskip0.2cm
       Max-Planck-Institut f\"ur Physik   \\
       -- Werner-Heisenberg-Institut -- \\
        D--80805, Munich (Fed. Rep. Germany)\\
\vskip0.5cm
 {\large H.~Simma}
\vskip0.2cm
    Deutsches Elektronen Synchrotron DESY \\
        D--22603 Hamburg (Fed. Rep. Germany)\\
\vskip1cm
{\Large Abstract\\}
\parbox[t]{\textwidth}{
  We carry out a detailed study of exclusive radiative rare $B$-decays
  in the framework of the QCD sum rules on the light cone, which
  combines the traditional QCD sum rule technique with the description
  of final state vector mesons in terms of the light-cone wave functions
  of increasing twist. Our calculation is restricted to the leading
  twist-two operators. The decays considered are: $B_{u,d} \to K^*
  + \gamma , ~B_{u,d} \to \rho + \gamma , ~B_d \to \omega + \gamma$
  and the corresponding decays of the $B_s$ mesons, $B_s \to \phi +
  \gamma$ and $B_s \to K^* + \gamma$.  Based on our estimate of
  the transition form factor $F_1^{B \to K^*\pg} (0)=0.32 \pm
  0.05$, we find for the branching ratio $BR(B \to K^* + \gamma)
  = (4.8\pm 1.5)\times 10^{-5}$, which
  is in agreement with the observed value of $(4.5\pm
  1.5\pm 0.9)\times 10^{-5}$ measured by the CLEO collaboration.  We
  present detailed estimates for the ratios of the radiative decay
  form factors, which are then used to predict the rates for the
  exclusive radiative $B$-decays listed above.  This in principle
  allows the extraction of the CKM matrix element $\vert V_{td} \vert$
   from
  the penguin-dominated CKM-suppressed radiative decays
  when they are measured.  We give a detailed discussion of the
  dependence of the radiative transition form factors on the $b$-quark
  mass and on the momentum transfer, as well as their interrelation
  with the CKM-suppressed semileptonic decay form factors in $B \to
  \rho + \ell + \nu_\ell$, which we also calculate in our approach.
}
\vskip2cm
{\em Submitted to Zeitschrift f\"ur Physik }
\end{center}
\noindent
CERN--TH.7118/93\\
December 1993
\newpage
\setcounter{page}{1}
% Decrease texheight (for preprint numbers) again
\textheight 23.0 true cm

%%%%%%%%%%%%%%%%%%%%%% SECTION 1 %%%%%%%%%%%%%%%%%%%%%%%%%%%%%%%%
\section{Introduction}
Theoretical interest in rare $B$-decays
 lies in the first place in their potential role
as precision tests of the quark sector of the Standard Model (SM).
Interpreted within this framework, the eventual measurements of these
decays will provide
quantitative information about the top quark mass and more
importantly about the Cabibbo-Kobayashi-Maskawa
 (CKM) matrix elements, $V_{td}, V_{ts}$ and
$V_{tb}$. In particular, the short-distance contribution to the
 CKM-suppressed rare $B$-decays $\BGAMAD$
and $b \to d + \ell^+ \ell^-$  directly
measures $V_{td}$. Together with improved measurements of the CKM matrix
elements $\Vcbabs$ and $\Vubabs$, this will
determine the CKM unitarity triangle resulting from the constraint
$\sum_{i}V_{id}V_{ib}^*=0$, and thus, pin down the CP-violating
phases in the Standard Model. These non-trivial constraints
on the CKM unitarity and the importance of rare $B$-decays in this
respect cannot be overemphasized. At the same time,
rare $B$-decays induced by flavour changing neutral currents (FCNC) have
the potential of providing one of the early hints for non-SM physics.
It is, therefore, imperative to get estimates of these FCNC
decays in the SM as reliable
 as possible and carry out an experimental physics
programme sensitive to rare $B$-decays.

\par
  The experimental searches for the FCNC $B$-decays has already
provided first dividends. The recent CLEO observation \cite{CLEO93}
  of the rare decay mode
$\BKSTAR$ having a combined branching ratio $\BBKSTAR =(4.5\pm 1.5 \pm
0.9)\times 10^{-5}$ and an improved upper limit on the inclusive
branching ratio $\BBGAMAXS < 5.4 \times 10^{-4} ~(95 \%$ C.L.)
\cite{THORND} have been analysed in the SM context,
as well as in several SM extensions.
Within the SM, the resulting experimental measurements have been
interpreted in terms of upper and lower limits on the CKM matrix
element ratio
$0.50 \leq \Vtsabs /\Vcbabs \leq 1.67$ (at $95 \%$ C.L.) \cite{ag5}.

\par
  On general grounds, the most quantitative tests of FCNC processes
in $B$-decays are the ones involving inclusive measurements such as
$B \to X_{d,s} + \gamma$ and the corresponding
semileptonic and purely leptonic processes such as $B \to X_{d,s} + \ell^+
\ell^-$ and $B_{d,s} \to \ell^+ \ell^-$.
Thus, the ratio of inclusive decay rates such as
$\BBGAMAXD /\BBGAMAXS$ would provide one of the cleanest determinations
of $\Vtdabs /\Vtsabs$ \cite{ag3}.
However, inclusice branching ratios are rather difficult to measure.
Exclusive decays such as $B \to \rho + \gamma , ~B  \to \omega + \gamma$
and the semileptonic decays $B \to (\pi, \rho, \omega) \ell^ + \ell^-$
are probably easier to measure, given a large number of $B$ hadrons and
a good electromagnetic detector and particle identification.
The interpretation of these and the already measured decay $B \to K^* +
\gamma$ necessarily needs reliable estimates of the decay form factors.
The decay $B \to K^* + \gamma $ has received quite a bit of theoretical
attention \cite{BKST1}--\cite{Santorelli}.

\par
The aim of this paper is to estimate a number of exclusive radiative
decays. These include the CKM-allowed decays, $B_{u,d} \to K^* +
\gamma$ and $B_s \to \phi + \gamma $, and the CKM-suppressed decays,
$B_{u,d} \to \rho + \gamma, ~B_d \to \omega + \gamma $ and
$B_s \to K^* + \gamma $.
While the effective Hamiltonian approach allows to include short-distance
QCD corrections at scales $\mu^2\ge m_b^2$,
estimates of the  hadronic matrix elements of the relevant operators
necessarily require some non-perturbative technique. We use here
QCD sum rules, which have proved to be a powerful tool for such
purposes, and in their classical form have been introduced in the
pioneering papers by Shifman, Vainshtein and Zakharov \cite{SVZ}.
In this paper, we use a modification of the QCD sum rule approach,
which we refer to as QCD sum rules on the light cone.
This technique has been
developed originally for light quark systems
in \cite{BBK,BF1}. The application for the heavy meson decays
was first suggested and studied in \cite{CZ-B}.

In this approach,
the ideas of duality and matching between the parton and hadron
descriptions, intrinsic to the QCD sum rule framework, are combined
with the specific techniques used in the studies of the hard exclusive
processes in QCD \cite{exclusive,BLreport}.
In contrast to the standard sum rule method for the evaluation
of form factors \cite{IS,NR}, in which the hadrons in the
initial and final state are treated in a symmetric way,
we resort to the QCD sum rule treatment of the initial $B$-meson
only, and describe the outgoing light vector meson by the set
of its  wave functions of increasing twist.
Hence the operator product expansion in the light-cone approach
 is governed  by
the twist of the operators rather than by their dimension, and the
vacuum expectation values of local operators are replaced by the
light-cone hadron wave functions.
 A physical motivation of the light-cone sum rule approach
in the present context
is an obvious asymmetry of the participating heavy initial
and light final state mesons.
The advantage of this formulation is that it allows to
incorporate additional information about high-energy asymptotics of
correlation functions in QCD, which is accumulated in
the wave functions.
The high energy behaviour of these functions is related to the
(approximate) conformal invariance of QCD, and
 many properties and results, following from this approximate
 invariance and
 obtained  previously, can be advantageously used in the
present context as well.
Existing applications of the light-cone sum rules include the
calculations of the amplitudes for the radiative decay $\Sigma\rightarrow
p\gamma $ \cite{BBK}, nucleon magnetic moments \cite{BF1}, the strong
couplings $g_{\pi N N}$ and $g_{\rho\omega\pi}$  \cite{BF1}, the
semileptonic $B$-meson \cite{CZ-B,BKR} and $D$-meson decay amplitudes
\cite{BBD}. In all these cases the results have been encouraging.

The principal theoretical result of this paper is a sum rule for the
electromagnetic penguin form factor appearing in the decays
 $B \to V + \gamma$, where $V$ is a vector meson. We derive the
corresponding sum rule in the traditional QCD sum rule approach also
  and comment on similar
existing sum rules in the literature \cite{Aliev}--\cite{Dom93}.  We
argue that the sum rules derived in the light-cone approach are more
reliable and we calculate the radiative transition form factors for a
number of exclusive radiative $B$-decays
in this framework. The main numerical results of our work can be
summarized as follows:
\begin{equation}
\begin{array}{rclrcl}
F_1^{B\rightarrow K^*\pg} &=& 0.32\pm 0.05 \hspace{2cm} &
F_1^{B\rightarrow (\rho,\omega)\pg} &=& 0.24\pm 0.04
\nonumber\\
F_1^{B_s\rightarrow \phi\pg} &=& 0.29\pm 0.05 &
F_1^{B_s\rightarrow K^*\pg} &=& 0.20\pm 0.04
\nonumber
\end{array}
\label{eq1}
\end{equation}
where the form factor $F_1$ is defined below (eq. (\ref{defF})).
The above estimate for the decay form factor
    $F_1^{B\rightarrow K^*\pg}$ can be combined with the QCD-improved
inclusive radiative branching ratio $BR(B \to X_s + \gamma)= (3.0 \pm
1.2)\times 10^{-4}$ \cite{ag5} to yield
$BR(B \to K^* + \gamma )=(4.8\pm 1.5)\times 10^{-5}$.
This agrees fairly well with the corresponding
branching ratio measured by the CLEO collaboration, posted as
$(4.5 \pm 1.5 \pm 0.9) \times  10^{-5}$ \cite{CLEO93}.
Together with the eventual measurements of the
 CKM-suppressed radiative decays $B_{u,d} \to \rho + \gamma ,
{}~B_d \to \omega + \gamma$ and $B_s \to K^* + \gamma$, the remaining
form factors in (\ref{eq1}) can then be used to determine the CKM
parameters.

While the main  thrust of the paper
is on providing theoretical estimates of the
exclusive radiative $B$-decay rates,
we also discuss a number of other, related, theoretical and
phenomenological issues. In particular, we
discuss in detail the heavy-quark-mass dependence of the
radiative $B$-decay form factors, and derive the sum rules
in the heavy quark limit.
%A comparison with the traditional
%QCD sum rules in this limit is also given and arguements
%in favour of our approach are presented.
A light-cone QCD sum rule for the
semileptonic decay $B \to \rho + \ell \nu_\ell$ is also derived
and the relation between radiative and semileptonic
form factors is studied.
\par
The paper is organized as follows. In section 2, we review the
effective Hamiltonian for radiative $B$-decays, dominated by the
electromagnetic penguins, and we work out explicitly the
dependence of their rates on the CKM matrix element parameters.
 A number of relations involving exclusive
and inclusive radiative $B$-decays is given in this section.
  In section 3, we describe
the method of the QCD sum rules on the light cone and derive our
general sum rule for the form factors of the electromagnetic penguin
operators entering in the decays $B \to V + \gamma$. The wave functions
of the vector mesons which are needed to evaluate the sum rule
 are discussed in section 4, and the numerical
estimates of the form factors are presented in section 5.
Section 6 includes the discussion of the $b$-quark mass
dependence of the form factors, and the heavy quark
limit. In section 7 we discuss the $q^2$ dependence and the
interrelation of radiative and semileptonic $B$-decay form factors
in the QCD sum rule approach.
Section 8 contains a summary and some concluding remarks, and
the impact of measuring the decays
$B_d\to \omega+\gamma, ~B_{u,d}\to \rho+\gamma$ and
$B_s\to K^*+\gamma$ in determining the CKM matrix element
$|V_{td}|$ is underlined.
The traditional three-point QCD sum rule for the radiative $B$-decay
$B \to K^* + \gamma$ is derived and discussed in the appendix.

%%%%%%%%%%%%%%%%%%%%%% SECTION 2 %%%%%%%%%%%%%%%%%%%%%%%%%%%%%%%%
\section{Effective Hamiltonian for inclusive and\newline
          exclusive radiative $B$-decays}
\par
In this section we review  the theoretical framework within the Standard
Model of electroweak interactions for inclusive and exclusive radiative
$B$-decays dominated by the electromagnetic penguins.
The inclusive decays
are grouped as $B \ra X_{f} + \gamma$, where we use the flavour of the
light quark $f=s,d$ in the transition $b \to f$ to characterize the
hadronic system recoiling against the photon.
Including lowest-order QCD corrections and gluon bremsstrahlung,
these decays are described at the parton level by the transitions
$b \to f \gamma$ and $b \to f \gamma g$. In calculating the
inclusive decay widths, we shall follow the work reported in \cite{ag5,
ag3, ag1}. The exclusive decays that
are the principal concern of this work are grouped in an analogous way
into $b\to s$ transitions:
\begin{itemize}
\item $B_u \to K^* + \gamma , ~~B_d \to K^* + \gamma$,
\item $B_s \to \phi + \gamma$,
\end{itemize}
which we shall also term CKM-allowed, according to the dominant
CKM matrix element dependence of their decay
rates, and $b\to d$ transitions:
\begin{itemize}
\item $B_d \to \rho + \gamma , ~~B_d \to \omega + \gamma, ~~B_u \to \rho
      + \gamma $,
\item $B_s \to K^* + \gamma$,
\end{itemize}
which will be called CKM-suppressed.

\par
The framework to incorporate (perturbative) short-distance QCD corrections
in a systematic way is that of an effective low energy theory with five
quarks. It is obtained by integrating out the  heavier degrees of freedom,
i.e. the top quark and $W^\pm$ bosons.
Since the running of $\alpha_s$ between $\mt$ (present estimates
$\mt=164 \pm 27$ GeV \cite{Swartz})  and $m_W$ is not very significant,
and since $m_W/m_t$ would not be a good expansion parameter,
it is a reasonable approximation to integrate out both the top quark and
the $W^\pm$ at the same scale.

Before using the unitarity properties of the CKM matrix, the
effective Hamiltonian relevant for the processes $b\to f\gamma$
and $b\to f\gamma g$ has the form
\begin{eqnarray}
H_{eff}^{(b \to f)} & = & - \frac{4 G_{F}}{\sqrt{2}} \, \left(
V_{tb}V_{tf}^* \, \sum_{j=3}^{8} C_{j}(\mu) \, O _{j}(\mu)
\right.\nonumber \\ &+& \left.
V_{cb}V_{cf}^* \,\sum_{j=1}^{8} C^\prime_{j}(\mu) \, O^\prime_{j}(\mu)
 +  V_{ub}V_{uf}^* \, \sum_{j=1}^{8} C^\pp_{j}(\mu) \, O^\pp _{j}(\mu)
\right)
\quad ,
\label{Hamiltonian}
\end{eqnarray}
where $G_F$ is the Fermi coupling constant and $j$ runs through a complete
set of operators with dimension up to six; the $C_{j}(\mu) $ are their
Wilson coefficients evaluated at the scale $\mu$.

We recall here that $O_1^{\prime(\prime)}$ and
$O_2^{\prime(\prime)}$ represent the colour-singlet
and colour-octet four-fermion operators, respectively, obtained from the
SM charged current Lagrangian written in the charge retention form;
in particular,
\begin{equation}
O^\prime_{2} = (\bar{c}_{L \alpha} \go{\mu} b_{L \alpha})
(\bar{f}_{L \beta} \gu{\mu} c_{L \beta})
\ \ \ {\rm and} \ \ \
O^\pp_{2} = (\bar{u}_{L \alpha} \go{\mu} b_{L \alpha})
(\bar{f}_{L \beta} \gu{\mu} u_{L \beta}) \ ,
\label{O2} \end{equation}
where $\alpha$ and $\beta$ are SU(3) colour indices.
The remaining operators are the same for all three CKM prefactors,
i.e.  $O^\prime_j = O^\pp_j = O_j$ for $j\not=1,2$.
At tree level, the only contribution to $b\to f\gamma$ comes from
the magnetic moment operator
\begin{equation}
O _{7} = \frac{e}{16\pi^2} \, \bar{f}_{} \, \sigma^{\mu
\nu} \, (m_b  R + m_f  L) \, b_{} \ F_{\mu\nu}\ ,
\label{O7} \end{equation}
where $L,R = (1\mp\gamma_5)/2$, and $e$ is the QED coupling constant.
The four-fermion operators $O _{3},...,O _{6}$ and the QCD magnetic
moment operator $O_8$, which is the gluonic counterpart of $O_7$, arise
from penguin diagrams in the full theory (before integrating out $W$ and
$t$). These operators enter indirectly in $b\to f \gamma$ decays due
to operator mixing and through (virtual and bremsstrahlung) gluon
corrections.

Taking into account the unitarity of the CKM matrix, only two terms in
the combinations $\lambda_i \equiv V_{ib} V^{\star}_{is}$ --- or in the
combinations $\xi_i \equiv V_{ib} V^{\star}_{id}$ in the case of
CKM-suppressed transitions --- are independent.
For  $b\to s$ transitions, one finds $\vert\lambda_u\vert
\ll \vert\lambda_{c}\vert,\vert\lambda_{t}\vert$;
 therefore, when neglecting terms proportional
to $\lambda_u$, the effective Hamiltonian (\ref{Hamiltonian}) can be brought
into a form proportional to $\lambda_t \approx - \lambda_c$:
\begin{equation}
H_{eff}^{(b \to s)}
       = - \frac{4 G_{F}}{\sqrt{2}} \, \lambda_{t} \, \sum_{j=1}^{8}
C_{j}(\mu) \, O_{j}(\mu) \quad .
\label{Hb2s} \end{equation}

For $b\to d$ transitions, where $\xi_u$, $\xi_c$ and $\xi_t$ are all of
the same order of magnitude, it is most convenient to choose $\xi_u$ and
$\xi_c$ as independent CKM factors during the matching for the Wilson
coefficients at $\mu = m_W$ and during the renormalization group evolution
to $\mu\approx m_b$. Since terms of order $O(m_u^2/m_W^2)$ and
$O(m_c^2/m_W^2)$ (or $O(m_u^2/m_t^2)$ and $O(m_c^2/m_t^2)$)
 are thereby neglected,
the terms proportional to $\xi_u$ and $\xi_c$ are multiplied by
just the same coefficient functions, i.e. $C^\prime_j(\mu) =
C^\pp_j(\mu)$ (although the corresponding operators are, of course,
different, see (\ref{O2})).
Finally, exploiting again $\xi_u + \xi_c = -\xi_t$, one ends up with
\begin{equation}
H_{eff}^{(b \to d)}
       = - \frac{4 G_{F}}{\sqrt{2}} \, \left(
\xi_t \, \sum_{j=3}^{8} C_{j}(\mu) \, O_{j}(\mu)
 - \sum_{j=1}^{2} C_{j}(\mu) \, \left\{
\xi_c O^\prime_{j}(\mu) + \xi_u O^\pp_{j}(\mu) \right\} \right)
\quad , \label{Hb2d} \end{equation}
where the Wilson coefficients $C_j(\mu)$ are precisely the same functions
as in (\ref{Hb2s}).
Details of the full operator basis, the matching of the Wilson coefficients
$C_j$ at $\mu\approx m_W$, and the complete leading-logarithmic
renormalization group evolution to $\mu \ll m_W$ can be found in the
literature \cite{bsall}.

\par
In estimates of the inclusive branching ratio for $\BBGAMAXS$ one has to
include the contribution of the QCD bremsstrahlung process \bsggam ~and
the virtual corrections to \bsgam , both calculated in $O(\alpha \alpha_s)$
in ref.~\cite{ag1}. Expressed in terms of the inclusive semileptonic
branching ratio $BR(B \to X \ell \nu_\ell)$, one finds:
\begin{equation}
  \BBGAMAXS = 6 \frac{\alpha}{\pi}
 \frac{|\lambda_{t}|^2}{|V_{cb}|^2}
 \frac{|C_7(x_t ,m_b)|^2 K(x_t, m_b)
                               }{g(m_c/m_b)( 1-2/3 \frac{\alpha_s}{\pi}
                f(m_c/m_b))}\times BR(B \to X \ell \nu_\ell)\ ,
    \label{e6}
\end{equation}
where $x_t=m_t^2/m_W^2$ indicates the explicit $m_t$-dependence in
$C_7(x_t ,m_b ) \equiv C_7(m_b )$, and
$ g(r)=1-8r^2 +8r^6-r^8-24r^4\ln (r)$
is the phase-space function for $\Gamma(b \to c + \ell \nu_\ell)$.
The function $f(r)$ accounts for QCD corrections to the semileptonic
decay and can be found,
 for example, in ref.~\cite{Alipiet}. It is a slowly
varying function of $r$, and for a typical quark mass ratio of
$r=0.35 \pm 0.05$, it has the value $f(r)=2.37 \mp 0.13$.
The contributions from the decays $b \to u + ~\ell ~\nu_\ell$
have been neglected in the denominator in (\ref{e6}) since they are
numerically  inessential ($\Vubabs\ll\Vcbabs$).
For the semileptonic branching ratio, the measured value
$BR(B \to X \ell \nu_\ell) \simeq 11\%$ will be used.

\par
             The inclusive decay width for $\BGAMAXS$
is dominantly contributed by the magnetic moment
 term $C_7(x_t,\mu) O_7(\mu)$, hence
the rationale of factoring out this coefficient in the expression for
$\BBGAMAXS$ in Eq. (\ref{e6}) above.
Including $O(\alpha_s)$ corrections
brings to the fore other operators with their
specific Wilson coefficients. The
effect of these additional terms can be expressed in terms of
the function $K(x_t, m_b)$, which is
a $K$-factor in the sense of QCD corrections.
The function $K(x_t, m_b)$ has been computed in \cite{ag5}
taking into account the dominant corrections from $C_2$ and $C_8$ (the
coefficients of other operators are considerably smaller \cite{bsall}).
The resulting branching ratio is\footnote{obtained by
setting the CKM matrix element ratio $\vert \lambda_t \vert ^2/\Vcbabs ^2$
to 1, according to our present knowledge of the CKM matrix
(readily seen, e.g.,
 in the Wolfenstein representation \cite{Wolfenstein}).}
\cite{ag5}:
\begin{equation}
\BBGAMAXS =(3.0 \pm 0.5) \times 10^{-4}
\end{equation}
for $\mt$  in the range  100~GeV $\leq ~\mt ~\leq $ 200~GeV.
This is to be contrasted with
the present upper limit on the exclusive radiative decay,
$\BBGAMAXS < 5.4 \times 10^{-4}$ (at 90\% C.L.) \cite{THORND}.

\par
It has been argued in ref. \cite{ag5} that
the scale dependence of $C_7(x_t, \mu)$ is very pronounced in
the presently available leading-logarithmic approximation.
This inherent uncertainty of the present theoretical framework has
to be borne in mind in a quantitative discussion of the QCD-improved decay
rates. In ref. \cite{ag5} this uncertainty has been estimated by varying
the scale parameter $\mu$ in the range $ m_b/2 \leq \mu \leq 2 m_b$ and
introduces an additional theoretical error of $\pm 1\times 10^{-4}$ in the
above estimates for $\BBGAMAXS$.

The case of inclusive $\BGAMAXD$ decays is somewhat more complicated
with respect to the factorization of the CKM parameters.
When evaluating the one-loop matrix elements for the
parton process $b\to s\gamma g$, at least the four-quark operators
$O^\prime_2$ and $O^\pp_2$ should be kept since their coefficients
are of order unity (they contribute in penguin diagrams emitting both
a gluon and a photon). These contributions involve
a non-trivial dependence on the CKM matrix elements through terms
proportional to $\xi_u$ and $\xi_c$, see eq.~(\ref{Hb2d}).
Explicit expressions for this dependence on the $\rho$ and $\eta$
parameters in the Wolfenstein representation of the CKM matrix have
been evaluated in ref.~\cite{ag3}.

\par
In the following we shall focus our attention on {\it exclusive}
decays, like $B_{u,d} \to K^* + \gamma$ and $B_s \to \phi + \gamma $, and
the corresponding CKM-suppressed modes listed at the beginning of this
section.
At tree level only the magnetic moment operator $O_7$ of
eq.~(\ref{O7}) contributes to the transition amplitudes.
For a generic radiative decay  $B \to V + \gamma$
one defines a transition form factor $F_1(q^2)$ as:
\begin{equation}
\langle V,\lambda |\bar{f} \sigma_{\mu\nu} q^\nu b
 |B\rangle
= i \epsilon_{\mu\nu\rho\sigma} e^{* (\lambda) \nu} p^\rho_B p^\sigma_V
2 F_1^{B\rightarrow V}(q^2)\ ,
\label{defF}
\end{equation}
where $V$ is a vector meson ($V=\rho, \omega, K^*$ or $\phi$)
with the polarization vector $e^{(\lambda)}$; and
$B$ is the generic $B$-meson $B_u, B_d$ or $B_s$.
The vectors $p_B$, $p_V$ and $q=p_B-p_V$
denote the four-momenta of the initial $B$-meson and the
outgoing vector
meson and photon, respectively.

In (\ref{defF}) it is understood that the operator is evaluated at
the scale $\mu =m_b$, and all large logarithms, $\ln(m_W/m_b)$ and
$\ln(m_t/m_b)$, are included in the coefficient function $C_7(\mu=m_b)$.
The $b$-quark mass, which has been factored out, should be identified
with the pole mass, although the complete two-loop treatment
of the coefficient function is needed to make this identification
meaningful.

In evaluating the hadronic matrix elements,
one may  consider the $b$-quark mass as a large parameter,
and try to collect logarithms,  corresponding to the
so-called hybrid anomalous dimension \cite{hybrid}.
At zero-recoil,  $q^2_{\rm max} = (m_B-m_{V})^2 \simeq 19$ GeV$^2$,
this treatment is simple and the answer is
obtained in the framework of the heavy quark effective theory.
However, in this case the extrapolation to the physical point
$q^2=0$ introduces a large uncertainty.
 We take a different approach,
and use the QCD sum rules to calculate the form factors directly
at the physical point $q^2=0$ and for the finite value of
the $b$-quark mass. The price to pay is that the treatment of
hybrid logarithms becomes complicated, and
we shall ignore them in this paper.

With the above definition, the exclusive decay widths are given
by ($B=B_u ~\mbox{or} ~B_d$):
\begin{equation}
\Gamma (\BGAMAKSTAR )=\frac{\alpha}{32 \pi^4} G_F^2
 \vert \lambda_t \vert^2 {\vert F_1^{B\rightarrow K^*}(0)}\vert^2
    C_7(x_t,m_b)^2 (m_b^2 +m_s^2)\frac{(m_B^2-m_{K^*}^2)^3}{m_B^3},
\label{e9}
\end{equation}
and the analogous expression for $\Gamma (B_s\to\phi+\gamma )$.
The branching ratios of these exclusive decays can again be written
in terms of the inclusive semileptonic branching ratio:
\begin{equation}
  \BBGAMAKSTAR = 6 \frac{\alpha}{\pi}
 \frac{|\lambda_{t}|^2}{|V_{cb}|^2}
 \frac{|C_7(x_t ,m_b)|^2 \vert F_1^{B \to K^*}(0) \vert^2}
  {g(m_c/m_b)( 1-2/3 \frac{\alpha_s}{\pi}f(m_c/m_b))}
 \frac{(1-{m_{K^*}}^2/m_B^2)^3}{(1-m_s^2/m_b^2)^3} \times (11\%).
    \label{e8}
\end{equation}

A good quantity to test the model dependence of the form factors
for the exclusive decay is the ratio of the exclusive-to-inclusive
radiative decay widths ($B =B_u$ or $B_d$):
\begin{equation}
R(K^*/X_s) \equiv \frac{\Gamma (B \to K^* + \gamma )}{
\Gamma (B\to X_s + \gamma )}
=\frac{(1-{m_{K^*}}^2/m_B^2)^3 }{(1-m_s^2/m_b^2)^3 }
 \frac{m_{B}^3}{m_b^3}
 \frac{{\vert F_1^{B_i\rightarrow K^*}(0)}\vert^2}{K(x_t, m_b)} \ .
\label{RKSXS1}
\end{equation}
Note that $K(x_t,m_b) \simeq 0.83$ is almost independent of $m_t$
(for the range $100 ~\mbox{GeV}
\leq m_t \leq 200 ~\mbox{GeV}$) \cite{ag5}.
The exclusive-to-inclusive ratio involving $B_s$-decays is defined in
an analogous way.

Since the same short-distance-corrected coefficient function $C_7(m_b)$
enters in the Hamiltonian for CKM-allowed (\ref{Hb2s}) and CKM-suppressed
(\ref{Hb2d}) modes, the QCD scaling is identical for the two-body decays
$\BGAMAS$ and $\BGAMAD$, and it does not affect the ratio of the decay widths
$\Gamma (b \to d + \gamma )/\Gamma (b \to s + \gamma )$.
The same applies for the exclusive decays such as
$B \to \rho + \gamma$ and
$B \to K^* + \gamma$, in which case
the CKM factors factorize in the decay amplitudes.

 From this observation a number of relations between the exclusive decay
rates follow in the Standard Model \cite{ag3}.
 This is exemplified by the decay
rates for $B_{u,d} \to \rho + \gamma$ and $B_{u,d} \to K^* + \gamma$:
\begin{equation}
\frac{\Gamma (B_{u,d} \to \rho + \gamma)}
     {\Gamma (B_{u,d} \to K^* + \gamma)} =
      \frac{\vert V_{td} \vert^2}{\vert V_{ts} \vert ^2}
      \frac{\vert F_1^{B \to \rho }(0)\vert^2}
          {\vert F_1^{B \to K^* }(0)\vert^2} \Phi_{u,d} \,,
\label{SMKR}
\end{equation}
where $\Phi_{u,d}$ is a phase-space factor:
\begin{equation}
\Phi_{u,d} = \frac{(m_b^2 + m_d^2)}{(m_b^2 + m_s^2)}
    \frac{(m_{B_{u,d}}^2 -m_\rho^2)^3}{(m_{B_{u,d}}^2 -m_{K^*}^2)^3}.
\end{equation}
The ratio (\ref{SMKR}) depends only on the CKM matrix elements
and the ratio of form factors, while it is independent of the
top quark mass (and of the renormalization scale $\mu$).

Note that the decay width
$\Gamma (B_d \to \omega + \gamma )$ is expected to be equal to the
decay width $\Gamma (B_d \to \rho + \gamma )$, apart from the minor
difference in the phase-space factors $\Phi$. This follows from the
assumption that the quark wave functions for $\omega$ and $\rho$
are described by the isoscalar and isovector  combinations,
 $\vert\omega\rangle= 1/\sqrt{2}(\bar{u}u + \bar{d}d)$
and $\vert\rho\rangle= 1/\sqrt{2}(\bar{u}u - \bar{d}d)$, respectively.

The CKM-suppressed exclusive decay width
$\Gamma (B_s \to K^* + \gamma )$
can be related to the CKM-allowed decay width
 $\Gamma (B_d \to K^* +\gamma )$ by a relation
similar to the one given in eq.~(\ref{SMKR}). However, one expects a
substantial difference in the form factors $F_1^{B_d \to K^*}(0)$
and $F_1^{B_s \to K^*}(0)$ due to the exchange of the roles of $s$ and
$d$ quarks in the wave function of the $K^*$ in the two decay modes.

%%%%%%%%%%%%%%%%%%%%%% SECTION 3 %%%%%%%%%%%%%%%%%%%%%%%%%%%%%%%%
\section{QCD sum rules on the light cone}
The aim of this and  the following two sections is to calculate
transition form factors, governing the radiative $B$-decays
 $B \to V + \gamma$, as  defined in (\ref{defF}).  Our approach
is very close to the calculation of the semileptonic
$B \to \pi e \nu $ form factor in \cite{CZ-B,BKR}.

To derive the sum rule, we consider the correlation function
\begin{equation}
 i \int dx\, e^{iqx}
\langle V(p,\lambda)| T\{
\bar \psi(x) \sigma_{\mu\nu} q^\nu b(x)
\bar b(0) i \gamma_5 \psi(0) \}|0\rangle
= i \epsilon_{\mu\nu\rho\sigma} e^{* (\lambda) \nu} q^\rho p^\sigma
 T((p+q)^2)
\label{correlator}
\end{equation}
at $q^2=0$, $p^2=m_V^2$,
 and at Euclidean $m_b^2-(p+q)^2$ of order several $\mbox{GeV}^2$.
Hereafter we use $\psi$ as a generic notation for the field of the
light quark.
Writing down the dispersion relation in $(p+q)^2$, we can separate
the contribution of the $B$-meson as the pole contribution
to the invariant function $ T((p+q)^2)$:
\begin{equation}
  T((p+q)^2)= \frac{f_B m_B^2}{m_b+m_q}\,\frac{2F_1(0)}{m_B^2-(p+q)^2}
+\ldots\,,
\label{pole}
\end{equation}
 where the dots stand  for contributions of higher-mass resonances
and the continuum. The $B$-meson decay constant is defined in
the usual way,
\begin{eqnarray}
    \langle 0| \bar \psi \gamma_\mu\gamma_5 b |B(p)\rangle =
 i p_\mu  f_B \,,
\end{eqnarray}
and $m_B$, $m_b$ and $m_q$
are the $B$-meson, $b$-quark and light-quark masses, respectively.

The virtuality of the heavy
quark in the correlation function under consideration is large, of order
$m_b^2-(p+q)^2$, and one can use the perturbative expansion
of its propagator in the external field of slowly varying
fluctuations inside the vector meson. The leading contribution
corresponds to the diagram shown in Fig. 1a, and equals
\begin{equation}
  \int dx\, e^{iqx} \int \frac{dk}{(2\pi)^4} \,e^{-ikx}
\frac{q_\nu}{m_b^2-k^2}
\langle V(p,\lambda)| T\{
\bar \psi(x) \sigma_{\mu\nu} (m_b+\not\!k)
i \gamma_5 \psi(0) \}|0\rangle .
\label{graph}
\end{equation}
To the leading-twist accuracy, taking into  account gluon corrections
like the one in Fig.~1b produces the path-ordered gauge factor in between
the remaining quark fields in (\ref{graph}), which we do not
show for brevity.

Thus, in general we are left with matrix elements of gauge-invariant
 non-local operators, sandwiched in between the vacuum
and the meson state.
These matrix elements define the light-cone
meson wave functions, which have received a lot of attention in
the past decade. Following Chernyak and
Zhitnitsky \cite{CZreport}, we define
\begin{equation}
\langle 0 |\bar\psi(0)\sigma_{\mu\nu}\psi(x)
|V(p,\lambda)\rangle =
i(e^{(\lambda)}_\mu p_\nu -e^{(\lambda)}_\nu p_\mu)
f_V^\perp \int_0^1 du\, e^{-iupx} \phi_\perp(u,\mu^2).
\label{twist2}
\end{equation}
Likewise,
\begin{eqnarray}
\langle 0 |\bar\psi(0)\gamma_\mu\psi(x)
|V(p,\lambda)\rangle &=& p_\mu \frac{(e^{(\lambda)} x)}{(px)}
f_V m_V\int_0^1 du\, e^{-iupx} \phi_\parallel(u,\mu^2)
\nonumber\\
 &&\mbox{}+\left( e^{(\lambda)}_\mu -p_\mu \frac{(e^{(\lambda)} x)}{(px)}
 \right) f_V m_V
 \int_0^1 du\, e^{-iupx} g_\perp^{(v)}(u,\mu^2),
\\
\langle 0 |\bar\psi(0)\gamma_\mu\gamma_5\psi(x)
|V(p,\lambda)\rangle &=&
-\frac{1}{4} \epsilon_{\mu\nu\rho\sigma} e^{(\lambda) \nu}
p^\rho x^\sigma  f_V m_V
\int_0^1 du\, e^{-iupx} g_\perp^{(a)}(u,\mu^2).
\label{twist3}
\end{eqnarray}
The functions $\phi_\perp(u,\mu^2)$ and $\phi_\parallel(u,\mu^2)$
give the leading-twist distributions in the fraction of total
momentum carried by the quark in transversely and longitudinally
polarized mesons, respectively. The functions
$g_\perp^{(v)}(u,\mu^2)$
and $g_\perp^{(a)}(u,\mu^2)$ are discussed in detail below.
The normalization is chosen  in such a
way that for all four distributions $f= \phi_\perp , \phi_\parallel ,
g_\perp^{(v)} , g_\perp^{(a)}$, we have:
 $$\int_0^1 du\, f(u)=1.$$
 In the matrix elements  of non-local
operators on the l.h.s. of (\ref{twist2})--(\ref{twist3}) the separations
 are assumed to be light-like, i.e. $x^2=0$. Regularization of
 UV divergences that arise in the process of the extraction of the
leading $x^2\rightarrow 0$ behaviour produces a non-trivial
scale dependence of the wave functions, which can be found by
renormalization group methods \cite{exclusive,BLreport}.
The scale in (\ref{graph}) is fixed by the actual light cone
separation, $\mu^2 \sim x^{-2} \sim m_b^2-(p+q)^2$.

Putting eqs. (\ref{graph})--(\ref{twist3}) together,
we obtain
\begin{eqnarray}
T((p+q)^2) &=& \int_0^1 du
\frac{1}{m_b^2+\bar u u m_V^2 -u(p+q)^2}
\Bigg[m_b f_V^\perp \phi_\perp(u) + u m_V f_V g_\perp^{(v)}(u)
\nonumber\\
&&\mbox{}
+\frac{1}{4} m_V f_V g_\perp^{(a)}\Bigg] +  \frac{1}{4}
 \int_0^1 du \frac{m_b^2-u^2m_V^2}{(m_b^2+\bar u u m_V^2 -u(p+q)^2)^2}
 m_V f_V g_\perp^{(a)}(u)  \,. \nonumber\\
\label{a111}
\end{eqnarray}
where $\bar u = 1-u$.
The expression in (\ref{a111}) has the form of a dispersion
integral in $(p+q)^2$, which can be made explicit by introducing
the squared mass of the intermediate state $s=m_b^2/u+\bar u m_V^2$
as the integration variable, instead of the Feynman
parameter $u$. The basic assumption of the QCD sum rule
approach is that the contribution of the $B$-meson
corresponds in this dispersion integral to  the contribution of
intermediate states with masses smaller than a certain
threshold $s< s_0$ (duality interval).
 Making the Borel transformation
$1/(s-(p+q)^2) \rightarrow \exp(-s/t)$ and equating the result to
the $B$-meson contribution in (\ref{pole}), we arrive at the
sum rule
\begin{eqnarray}
\lefteqn{\frac{f_B m_B^2}{m_b+m_q} 2 F_1(0) e^{-(m_B^2-m_b^2)/t}  =}
\nonumber\\ &=&
 \int_0^1 du \frac{1}{u}\exp\left[-\frac{\bar u}{t}
\left(\frac{m_b^2}{u} + m_V^2\right)\right]
\theta\left[s_0-\frac{m_b^2}{u}-\bar u m_V^2\right]
\Bigg\{m_b f^\perp_V \phi_\perp(u,\mu^2=t)
\nonumber\\
&&\mbox{}+
 u m_V f_V g_\perp^{(v)}(u,\mu^2=t)
 +\frac{m_b^2-u^2m_V^2+u t}{ 4 u t} m_V f_V
 g_\perp^{(a)}(u,\mu^2=t) \Bigg\} \,,
\label{SR}
\end{eqnarray}
which should be satisfied for values of the Borel parameter $t$
of order several $\mbox{GeV}^2$.  To the leading logarithmic accuracy,
the scale in the wave functions coincides with the
Borel parameter.
Principal input in this sum rule are the vector meson wave functions,
which contain non-trivial information about the dynamics at large
distances, and which we are going to discuss now.

%%%%%%%%%%%%%%%%%%%%%% SECTION 4 %%%%%%%%%%%%%%%%%%%%%%%%%%%%%%%%
\section{Wave functions of the vector mesons}

It is known that the decomposition of the leading-twist meson wave
functions in terms of conformal invariant operators allows one to
diagonalize the mixing matrix at one-loop order. Thus,
the (approximate) conformal invariance of QCD implies that
the coefficients in the expansion of leading-twist wave functions
in the series of Gegenbauer polynomials \cite{BE} are renormalized
multiplicatively to the leading logarithmic accuracy.
In our case
\begin{eqnarray}
\phi_\perp(u,\mu)&=&6 u (1-u) \left[1 +
a_1(\mu) \xi + a_2(\mu)  \left(\xi^2-\frac{1}{5}\right)
  +  a_3(\mu)
 \left(\frac{7}{3}\xi^3-\xi\right) + \ldots\right]\,,
 \nonumber\\
  a_n(\mu) &=& a_n(\mu_0)
\left(\frac{\alpha_s(\mu)}{\alpha_s(\mu_0)}\right)^{\gamma_n/b}\,,
\label{wf1}
 \end{eqnarray}
where we have introduced the shorthand notation $\xi=2u-1$.
Here $b=(11/3) N_c - (2/3) n_f$.
 The anomalous dimensions turn out to be \cite{SV}
\begin{equation}
\gamma_n = C_F \Bigg( 1+4 \sum_{j=2}^{n+1} 1/j \Bigg) ,
 \end{equation}
where $C_F = (N_c^2-1)/(2N_c)$.
The coefficients in the Gegenbauer expansion  (\ref{wf1})
at a low scale, $a_n(\mu_0)$,
 should be determined by a certain non-perturbative
approach, or taken from experiment. At present, most of the
existing information comes from the QCD sum rules. Following
\cite{CZreport}, we use the model wave functions
for $\rho$, $K^*$ and $\phi$ mesons at the scale $\mu_0^2=1$ GeV$^2$,
 corresponding to the following choice of the parameters in
 (\ref{wf1}):
\begin{eqnarray}
  & a_2^{(\rho)} = -1.25, &
\nonumber\\
a_1^{(K^*)} = 0.75,  & a_2^{(K^*)} = -2, & a_3^{(K^*)} = 0.75,
\nonumber\\
 & a_2^{(\phi)} = -2.5  . &
\label{low}
\end{eqnarray}
Note that in the case of $\rho$ and $\phi$ mesons only $a_2$ is
non-zero, and the wave function
is symmetric under the interchange $u\leftrightarrow 1-u$.
In the case of a $K^*$-meson, it is necessary to specify that
the variable $u$ is the fraction of the total momentum carried by
the strange quark.
The decay constants appearing in (\ref{twist2}) are \cite{CZreport}:
\begin{eqnarray}
   f^\perp_\rho &=& 200\,\mbox{\rm MeV}\,,
\nonumber\\
   f^\perp_{K^*}&=& 210\,\mbox{\rm MeV}\,,
\nonumber\\
   f^\perp_\phi&=& 230\, \mbox{\rm MeV}\,.
\label{fperp}
\end{eqnarray}
As mentioned above, the correct
normalization point in the sum rules is of the order of the
typical Borel parameter, which for $B$-meson decays
is of order $\mu^2\sim m_B^2-m_b^2 \sim 5$ GeV$^2$. Using
$\Lambda_{\overline{MS}}= 225$ MeV in
the renormalization group  rescaling factors given in (\ref{wf1}),
we obtain at this scale
\begin{eqnarray}
  & a_2^{(\rho)} = -0.85, &
\nonumber\\
a_1^{(K^*)} = 0.57,  & a_2^{(K^*)} = -1.35, & a_3^{(K^*)} = 0.46,
\nonumber\\
 & a_2^{(\phi)} = -1.7 . &
\label{high}
\end{eqnarray}
The wave functions corresponding to the above parametrization
are shown in Fig.~2.
As a general effect of the rescaling, the wave functions
become somewhat wider and  closer to the asymptotic
expression $6 u(1-u)$.
In the case of the $K^*$-meson, the wave function  becomes  both
wider and more symmetric, so that scaling violation affects the region
$u<1/2$ only ( see Fig. 2b). The integration over $u$ in the
sum rule in (\ref{SR}) is restricted to the interval
of rather large momentum fractions, carried by the light quark involved
in the electromagnetic penguin operator.
For realistic values of parameters, see below, one has $u>0.65$--$0.7$.
According to the analysis in \cite{CZreport}, the $K^*$ wave function
turns out to be in this region very close to the asymptotic expression
$6 u(1-u)$. In the case of the decay $B_s\rightarrow K^*\gamma$, the
role of the strange and non-strange quarks in the $K^*$-meson
is reversed, corresponding to the formal substitution
$u\rightarrow 1-u$ in the wave function. Thus, the
region of small values $u<0.3-0.35$ in Fig. 2b becomes
appropriate, where
deviations from the asymptotic form are large and the effect
of rescaling is important. This results in a significant
difference between the form factors in the decays $B_d \to K^* + \gamma$
and $B_s \to K^* + \gamma$.

The conformal expansion of the wave functions
 $g_\perp^{(v)}(u,\mu^2)$
 and $g_\perp^{(a)}(u,\mu^2)$ is somewhat more involved and can be
obtained using the approach of \cite{BF2}.
To this end we define new wave functions, which contain quarks
with fixed spin projections on the light cone:
\begin{eqnarray}
\langle 0|\bar\psi(0)\not\!x\gamma_\mu\not\!p\psi(x)
|V(p,\lambda)\rangle &=&
- e^\lambda_\mu m_V f_V  (px)
\int_0^1 du\, e^{-ipxu} g^{\ud}(u,\mu^2),
\nonumber\\
\langle 0|\bar\psi(0)\not\!p\gamma_\mu\not\!x \psi(x)
|V(p,\lambda)\rangle &=&
- e^\lambda_\mu m_V f_V (px)
\int_0^1 du\, e^{-ipxu } g^{\du}(u,\mu^2).
\label{spin}
\end{eqnarray}
The wave functions $g^{\ud}(u)$ and $g^{\du}(u)$ can be expanded
in terms of irreducible representations of the collinear conformal group
as \cite{M,O,BF2}:
\begin{eqnarray}
  g^{\ud}(u)&=&2(1-u)\left[
 1 + c^{\ud}_1 P_1^{(1,0)}(\xi) +c^{\ud}_2 P_2^{(1,0)}(\xi) +\ldots
\right]\,,
\nonumber\\
  g^{\du}(u)&=&2u\left[
 1 + c^{\du}_1 P_1^{(0,1)}(\xi) +c^{\du}_2 P_2^{(0,1)}(\xi) +\ldots
\right]\,,
\label{spin1}
\end{eqnarray}
where $P_k^{(l,m)}(\xi)$ are Jacobi polynomials \cite{BE}.
The normalization in (\ref{spin}) follows from the standard
definition of vector decay constants
\begin{equation}
 \langle 0|\bar\psi\gamma_\mu\psi|V(p,\lambda)\rangle
=e^\lambda_\mu m_V f_V.
\label{fV}
\end{equation}
The numerical values determined from QCD sum rules
are \cite{SVZ,CZreport}
\begin{eqnarray}
   f_\rho &=& 200\,\mbox{\rm MeV}\,,
\nonumber\\
   f_{K^*}&=& 210 \,\mbox{\rm MeV}\,,
\nonumber\\
   f_\phi&=& 230\,\mbox{\rm MeV}\,.
\label{fvect}
\end{eqnarray}
Note that these couplings,
 to the accuracy of the existing QCD sum rule calculations,
coincide with the couplings in (\ref{fperp}).

Neglecting $SU(3)$-breaking effects  related to the difference
 of the quark and antiquark masses in the mesons,
one has
\begin{equation}
       c^{\ud}_k = (-1)^k c^{\du}_k  \equiv c_k .
\label{symmetry}
\end{equation}
Finally, the coefficient $c_1$ is actually fixed by the equations of
motion, which allow one to reduce the matrix element
\begin{equation}
 \langle 0|\bar\psi\gamma_\mu\gamma_5
 (i\stackrel{\leftrightarrow}{D}_\nu)\psi|V(p,\lambda) \rangle
= i A \epsilon_{\mu\nu\rho\sigma} e^\rho p^\sigma
\end{equation}
to the matrix element in (\ref{fV}).
Thus, we find $ A=-(1/2) f_V m_V $ \cite{ZZC} and
\begin{equation}
                 c_1 = -1/2\,.
\end{equation}

On the other hand, one has the obvious relations:
\begin{eqnarray}
g_\perp^{(v)}(u)&=&\frac{1}{2}[g^{\ud}(u)+g^{\ud}(u)]\,,
\nonumber\\
\frac{d}{du}g_\perp^{(a)}(u)&=&2[g^{\ud}(u)-g^{\ud}(u)]\,.
\label{wf2}
\end{eqnarray}
Using the identities \cite{BE}
\begin{eqnarray}
(1+\xi) P^{(0,1)}_k(\xi)
+(1-\xi) P^{(1,0)}_k(\xi) = 2C_k^{1/2}(\xi)\,,
\nonumber\\
(1+\xi) P^{(0,1)}_k(\xi)
-(1-\xi) P^{(1,0)}_k(\xi) = 2C_{(k+1)}^{1/2}(\xi)\,,
\nonumber\\
\frac{d}{d\xi} (1-\xi^2) C^{3/2}_k(\xi) =
-(k+1)(k+2) C_{k+1}^{1/2}(\xi)\,,
\end{eqnarray}
where $C_k^\lambda(\xi)$ are Gegenbauer polynomials,
we arrive at the expansions:
\begin{eqnarray}
  g_\perp^{(v)}(u) &=& \sum_{k=2n} (c_k-c_{k-1}) C^{1/2}_k(\xi)\,,
\nonumber\\
 g_\perp^{(a)}(u) &=& 2 (1-\xi^2)
 \sum_{k=2n} \frac{c_k-c_{k+1}}{(k+1)(k+2)}C^{3/2}_k(\xi)\,.
\label{expand}
\end{eqnarray}
The coefficients  in front of each Gegenbauer polynomial
come from the operators with two neighbouring conformal
spins. According to a general result \cite{M}, the
operators with different conformal spin do not mix under
renormalization to leading logarithmic accuracy. This is a major
simplification but does not guarantee multiplicative renormalization
in the present case because of the existence of three-particle
antiquark--quark--gluon operators with the same twist and conformal
spin (see  \cite{BF2}).

Further insight in the structure of the wave functions
$g_\perp^{(v)}(u)$ and $g_\perp^{(a)}(u)$ can be obtained by
using the equations of motion. Note that both these wave functions
correspond to transverse spin distributions, and involve precisely
the same operators
(apart from a missing $\gamma_5$ in the case of $g_\perp^{(v)}(u)$)
as the ones involved in the operator product expansion for the
structure function $g_2(x,Q^2)$ of
polarized deep inelastic lepton--nucleon scattering.
The difference is indeed in the matrix elements -- which involve the
(polarized) nucleons with equal momenta in the latter case and
the vacuum and the vector meson state in the present
situation. Thus, similar to the case of the structure function
$g_2$, the wave functions $g_\perp^{(v)}(u)$ and $g_\perp^{(a)}(u)$
contain contributions coming from both operators of twist 2 and
twist 3. In close analogy to deep inelastic scattering,
the twist-2 contributions to the ``transverse"
wave functions $g_\perp^{(v)}$,
$g_\perp^{(a)}$ can be expressed in terms of
the leading-twist ``longitudinal"
wave function $\phi_\parallel(u,\mu^2)$ defined in (\ref{twist3}):
\begin{eqnarray}
 g_\perp^{(v),{\rm twist-2}}(u) &=&
\frac{1}{2}\left[
 \int_0^u dv \frac{\phi_\parallel(v)}{\bar v}
               + \int_u^1 dv \frac{\phi_\parallel(v)}{ v}\right]\,
\nonumber\\
    \frac{d}{du}g_\perp^{(a),{\rm twist-2}}(u) &=&
2\left[  -
 \int_0^u dv \frac{\phi_\parallel(v)}{\bar v}
               + \int_u^1 dv \frac{\phi_\parallel(v)}{ v}\right]\,
\label{WW}
\end{eqnarray}
which is the analogue of the Wandzura-Wilczek relations \cite{WW}
between the structure functions $g_2$ and $g_1$.
The remaining twist-3 contributions to $g_\perp^{(v)}$, $g_\perp^{(a)}$
can be written in terms of three-particle antiquark--gluon--quark
wave functions of transversely polarized vector mesons, considered in
\cite{ZZC,CZreport}. The derivation of (\ref{WW}) and the relations
for twist-3 contributions will be given elsewhere.

In this paper, we do not take into account contributions
of twist 3 which come from the gluon-exchange diagram
in Fig. 1b, and to this accuracy we need to keep the twist-2
contributions to the wave functions
$g_\perp^{(v)}$, $g_\perp^{(a)}$ only. In anology with a
similar situation in the case of the structure function $g_2$, one should
expect that corrections to the asymptotic wave function
$\phi_\parallel(u) = 6 u (1-u)$,
coming from twist-2 operators involving gluons \cite{CZreport}
are of the same order of magnitude as neglected contributions
of twist 3. Thus, for consistency, we must neglect the gluon
corrections to  $g_\perp^{(v)}$, $g_\perp^{(a)}$ altogether, which
amounts to putting to zero all the coefficients in (\ref{spin1})
except the first two, $c_0=1$ and $c_1=-1/2$.
Thus, we obtain, finally
\begin{eqnarray}
  g_\perp^{(v)}(u) &=& 1+\frac{1}{2} C^{1/2}_2(\xi) =
     \frac{3}{4}(1+\xi^2)\,,
\nonumber\\
 g_\perp^{(a)}(u) &=& \frac{3}{2} (1-\xi^2) = 6 u (1-u) \,,
\label{otvet}
\end{eqnarray}
 which we use in the numerical analysis.
It is easy to check that these expressions are exactly the ones
that follow from (\ref{WW}) with the asymptotic expression for
the longitudinal distribution amplitude $\phi_\parallel(u) =
6 u (1-u) $.   The  same expressions for the
transverse wave functions in (\ref{otvet})
have been found in \cite{ZZC} by a
different method. The identification of these contributions
as being of twist 3 in ref. \cite{ZZC} is, however, an error.

%%%%%%%%%%%%%%%%%%%%%% SECTION 5 %%%%%%%%%%%%%%%%%%%%%%%%%%%%%%%%
\section{Evaluation of the sum rules}

To complete the list of entries which appear in the sum rule
(\ref{SR}) we must specify the value of the pole $b$-quark mass
$m_b$ and the continuum threshold $s_0$. We vary these parameters
in the limits
\begin{eqnarray}
m_b&=&4.6 \mbox{--} 4.8 \, \mbox{GeV} \ ,
\nonumber\\
s_0^{(B)} &= & 33 \mbox{--} 35 \, \mbox{GeV}^2 \ ,
\label{limits}
\end{eqnarray}
and for the continuum threshold in the $B_s$-channel, we
take
\begin{equation}
   s_0^{B_s}-s_0^{B} = m_{B_s}^2 - m_B^2 \simeq 1\,~\mbox{GeV}^2 ,
\label{s0Bs}
\end{equation}
making use of the present world average $m_{B_s}=5373.2 \pm 4.2 \,\mbox{MeV}$
\cite{ALEPH,Danilov,Martinerice}.

The value of the $B$-meson decay constant $f_B$ has received a lot
of attention recently. Within the QCD sum rule approach, its value
is strongly correlated with the values of the $b$-quark mass and
the continuum threshold (see \cite{D} and references cited therein).
For consistency, we substitute the value of $f_B$ in (\ref{SR})
by the square root of the corresponding QCD sum rule:
\begin{eqnarray}
\lefteqn{\frac{f_B^2 m_B^4}{m_b^2} e^{-(m_B^2-m_b^2)/t} }
\nonumber\\
&=&
\hspace{-4mm}
\frac{3}{8\pi^2}\int_{m_b^2}^{s_0}s ds\,e^{-(s-m_b^2)/t}(1-m_b^2/s)^2
 -m_b\langle \bar q q\rangle_{\mu^2=t}
- m_b\langle \bar q\sigma \mbox{\rm g}G q\rangle_{\mu^2=t}
\frac{1}{2t}\left(1-\frac{m_b^2}{2t}\right) \,,\nonumber\\
\label{SRB}
\end{eqnarray}
in which we discarded the radiative $O(\alpha_s)$ corrections,
as they are not taken into account in the sum rule (\ref{SR}) either,
and we also neglected numerically insignificant
 contributions of the four-quark operators
and of the gluon condensate.\footnote{
The sum rule in (\ref{SRB}) yields  $f_B\sim 120$--$160$ MeV,
which is lower than the value preferred at present,
typically $f_B\sim 180$ MeV (see \cite{D} for a review).
This smaller value is an artefact of neglecting
the radiative corrections, which
are numerically large. One can hope that these corrections
will be reduced significantly by taking the ratio of the
sum rules. By  direct calculations
 in the limit $m_b\rightarrow\infty$, it has been shown on
examples of increasing complexity (the Isgur-Wise function
\cite{BBG,Neubert}
and  the kinetic energy operator \cite{BB}) that
radiative corrections are indeed
very strongly reduced by considering
 the sum rule  ratios, and we conform to this practice here.}
 The sum rule in (\ref{SRB}) is evaluated
at precisely the same values of $m_b$ and $s_0$ as in (\ref{SR}),
and using the standard values of the vacuum condensates
$\langle \bar q q\rangle = -(240$ MeV$)^3$
 and $\langle\bar q\sigma\mbox{\rm g}G q\rangle/\langle \bar q q\rangle
 =0.8$ GeV$^2$ (at the low scale $\mu_0^2=1$ GeV$^2$), which have been
used in the QCD sum rule analysis of the wave functions in \cite{CZreport}.
The renormalization group evolution is given by
\begin{eqnarray}
 \langle \bar q q\rangle_{\mu^2} &= &
\left(\frac{\alpha_s(\mu)}{\alpha_s(\mu_0)}\right)^{-4/b}
\langle \bar q q\rangle_{\mu_0^2}\,,
\nonumber\\
 \langle \bar q \sigma  \mbox{\rm g}G q\rangle_{\mu^2} &= &
\left(\frac{\alpha_s(\mu)}{\alpha_s(\mu_0)}\right)^{2/(3b)}
\langle \bar q \sigma \mbox{\rm g}G q\rangle_{\mu_0^2}\,.
\end{eqnarray}
For the decay constant of the $B_s$-meson, we use the sum rule
in (\ref{SRB}), and the  ratio
\begin{equation}
     f_{B_s}/f_{B_d} =1.15\pm 0.05,
\end{equation}
which is in agreement with most existing estimates \cite{Dom93,FBS}.

The stability plots for the form factors as functions
of the Borel parameter are shown in Fig. 3. The stability is
in all cases good, and the variations of the
 continuum threshold within the limits (\ref{limits})
induce uncertainties in the values of form factors within
 less than 3\%.
Using the region in the Borel parameter
$6 ~\mbox{GeV}^2 <t<9 \, ~\mbox{GeV}^2$,
 we extract the values of form factors
\begin{eqnarray}
     F_1^{B\rightarrow K^*\pg} &=& 0.32 \pm 0.05
\nonumber\\
    F_1^{B\rightarrow (\rho,\omega) \pg} &=& 0.24\pm 0.04
\nonumber\\
    F_1^{B_s\rightarrow \phi\pg} &=& 0.29\pm 0.05
\nonumber\\
     F_1^{B_s\rightarrow K^*\pg} &=& 0.20\pm 0.04
\label{ffactors}
\end{eqnarray}
The given error mainly
comes from the uncertainty in the $b$-quark mass
and the dependence on the Borel parameter. In addition, we have
included an uncertainty of 20\% for the parameters (\ref{low})
of the wave functions. This leads for all decays to an error of order
$\pm(0.01$--$0.015)$ for $F_1$.
For all decays except $B_s\to K^*\gamma$, the sum rule is dominated
by the contribution of the wave function $\phi_\perp$, and the average
value of the momentum fraction $u$ under the integral in
(\ref{SR}) is $\langle u\rangle \simeq 0.8$.

The uncertainties in the values of input parameters
tend to be reduced in the form factor ratios, see Fig. 4,
and we obtain
\begin{equation}
\frac{ F_1^{B\rightarrow (\rho,\omega) \pg}}
      { F_1^{B\rightarrow K^*\pg}} =0.76\pm 0.06.
\label{ratios}
\end{equation}
which is not sensitive to the value of the $B$-meson decay constant.
Similarly, we have
\begin{equation}
\frac{ F_1^{B_s\rightarrow K^*\pg}}
      { F_1^{B_s\rightarrow \phi\pg}} =0.66\pm 0.09
\label{ratiobs}
\end{equation}
and
\begin{equation}
\frac{ F_1^{B_s\rightarrow K^*\pg}}
      { F_1^{B\rightarrow K^*\pg}} =0.60\pm 0.12.
\label{ratioks}
\end{equation}
The errors given for these ratios are dominated by uncertainties
in the $SU(3)$-breaking in the wave functions, while the quark-mass
dependence is greatly reduced.
The small value of the last ratio is due to the asymmetry of
the $K^*$ wave function -- the strange quark carries a larger
fraction of the meson momentum (see Fig. 2).    It should be
noted that errors given here
reflect uncertainties in the input parameters,
but do not include a possible ``theoretical" uncertainty of
the method itself. We estimate the actual accuracy
of the calculation to be within 20--30\% for the form factors and
10--20\% for their ratios.
 To improve the accuracy, it is necessary to
calculate corrections to the sum rule (\ref{SR}) coming
from antiquark--gluon--quark components of the vector meson wave
functions, $SU(3)$-breaking corrections to the wave functions
$g_\perp^{(a,v)}$,
and the perturbative QCD radiative corrections.

%%%%%%%%%%%%%%%%%%%%%% SECTION 6 %%%%%%%%%%%%%%%%%%%%%%%%%%%%%%%%
\section{The $\bf b$-quark mass dependence and\newline the heavy quark limit}

In order to facilitate the
comparison of  the QCD sum rule results with lattice
calculations \cite{BHS93,UKQCD},
we have done the numerical analysis of the sum rules
(\ref{SR}) for an arbitrary value of the $b$-quark mass, assuming
the following parametrization
suggested by the heavy quark effective theory \cite{HQETreview}:
\begin{eqnarray}
      m_B &=& m_b+\bar\Lambda \,,\hspace{1cm}
\bar\Lambda \simeq 500\mbox{--}700 ~\mbox{\rm MeV}\,,
\nonumber\\
      s_0^B - m_b^2 &=& 2m_b\omega_0  \,,\hspace{1cm}
\omega_0 \simeq 1.0\mbox{--}1.2 ~\mbox{\rm GeV}\,,
\nonumber\\
t&=&2 m_b \tau \,,\hspace{1cm} \tau =0.5\mbox{--}0.8 ~\mbox{\rm GeV}\,.
\label{hqetscaling}
\end{eqnarray}
Here $\bar \Lambda$ is the binding energy in the limit
$m_b\rightarrow\infty$, and $\omega_0$ is the
continuum threshold.
 The variable $\tau$ has the meaning of the non-relativistic
Borel parameter \cite{S}--\cite{BBBD}.

The results are shown in Fig. 5, where for the reason which is explained
below we have plotted the value of the form factor multiplied by
the meson mass to the power $3/2$.
 In addition to the curves describing the quark-mass dependence in
the heavy-quark parametrization (\ref{hqetscaling}), we also show our
result for $F_1^{B\to K^*\pg}$ given in (\ref{ffactors}).
Furthermore, using the mass of the $D$-meson, instead of $m_B$, and the
standard values of the parameters for $D$-mesons, $s_0^D=6$ GeV$^2$,
$m_c=1.4\pm0.05$ GeV, and $t=1.5$--3 GeV$^2$ \cite{BBD}, we obtain
\begin{equation}
 F_1^{D\to K^*}=0.85\pm 0.13 \ .
\label{F1D}
\end{equation}
Although the number given in (\ref{F1D}) is not
directly physically relevant, it can facilitate the comparison
with the present lattice data, which  are collected mostly
 in the $D$-meson mass range.

The curves in Fig. 5 disagree slightly with our values for the
form factors for $B$ and $D$-mesons. For $B$-mesons, the reason is that
in (\ref{limits}) we use a rather generous uncertainty in the
$b$-quark mass. For $D$ mesons, the omitted $O(1/m)$ corrections
to the parameters given in (\ref{hqetscaling}) can already become
significant.

 Comparing the values given in (\ref{ffactors}) and (\ref{F1D}),
we conclude that in the region between the charm and beauty
masses the form factor scales approximately as  $F_1 \sim 1/m_B$.
However, this power behaviour holds only for the sum rule
results in the constrained interval of quark masses, and
should not be confused with the theoretical behaviour
in the $m_b\rightarrow\infty$ limit, which we
are going to discuss now.

In the heavy $b$-quark limit the sum rule in (\ref{SR}) involves
the integration over a small interval of  $u$, of
order $1-u \sim 1/m_b$. This condition has a simple interpretation:
the spectator antiquark must recombine with the fast quark coming from
the $b$-decay. Therefore, the form factor is determined by the ``tail"
of the vector meson wave function corresponding to a strongly asymmetric
configuration where almost all the momentum is carried by one
of the constituents. It has been proven \cite{exclusive,BLreport} that
the behaviour of the wave functions in this region at sufficiently large
 virtualities
must coincide with the perturbative behaviour, which can be obtained
from the conformal properties of the corresponding operators.
Quite rigorously, one obtains for the wave functions involved in (\ref{SR}):
$\phi_\perp(u) \sim \phi_\parallel(u) \sim g_\perp^{(a)}\sim O(1-u)$,
and $g_\perp^{(v)}\sim O(1)$ for $1-u\rightarrow 0 $ (see
(\ref{wf1}) and (\ref{expand})).
It is easy to check that this behaviour is exactly the one
needed to ensure that the three terms in (\ref{SR}) are
of the same order in the limit $m_b\rightarrow\infty$, yielding
the form factor $F_1(0) \sim O(m_b^{-3/2})$. The diagram with the
hard gluon exchange produces the same power
behaviour \cite{CZ-B,Burdman},
and is estimated to be small \cite{Burdman}.
The large factors $\sim m_b$ in front of the wave functions
$\phi_\perp$ and $g_\perp^{(a)}$ in (\ref{SR})
 are compensated by the smallness
of the wave functions $\sim 1/m_b$ in the relevant region
$1-u\sim 1/m_b$.

In order to establish a connection to the heavy-quark expansion of
the two-point sum rules \cite{S}--\cite{BBBD}, we introduce
the scale-invariant (up to logarithms) form factor and $B$-decay constant
\begin{eqnarray}
       \widehat F_1(0)&=& F_1(0) \cdot m_B^{3/2} \,,
\nonumber\\
       \widehat f_B &=& f_B \cdot \sqrt{m_B}\,,
\end{eqnarray}
and  rewrite the sum rule (\ref{SR}) in terms of the
non-relativistic Borel parameter $\tau = t/(2m_b)$.
The result reads as:
\begin{equation}
\widehat f_B  \widehat F_1(0) e^{-\bar \Lambda/\tau}=  f_V
\int_0^{\omega_0} d\omega\, e^{-\omega/\tau}
\left\{-2\omega \phi'_\perp(1) +m_V g_\perp^{(v)}(1)-
\frac{\omega}{4\tau} m_V (g_\perp^{(a)})'(1)\right\}\,,
\label{SRHQL}
\end{equation}
where the new parameters
$\bar \Lambda $ and
 $\omega_0$  are defined in (\ref{hqetscaling}).
 The variable $\omega = m_b \bar u/2$ has the meaning
of frequency, and is defined as $(p+q)^2-m_b^2 = 2 m_b \omega$ (see
eq. (\ref{a111})). For the wave functions, we obtain, to our accuracy:
\begin{eqnarray}
-\phi'_\perp(1) \equiv -\frac{d}{du}\phi_\perp(1) &=&
 6[1+a_1(\mu) + (4/5) a_2(\mu) + (4/3) a_3(\mu) ] \,,
\nonumber\\
 g_\perp^{(v)}(1) &=& 3/2\,,
\nonumber\\
 -(g_\perp^{(a)})'(1) &=& 6\,,
\end{eqnarray}
where $\mu\simeq 2 \tau$ \cite{BBBD}.
In the same limit, the sum rule for the $B$-decay constant becomes
\cite{S}--\cite{BBBD}:
\begin{equation}
\widehat f_B^2  e^{-\bar \Lambda/\tau}=  \frac{3}{\pi^2}
\int_0^{\omega_0} d\omega\, e^{-\omega/\tau} \omega^2
 -\langle \bar q q\rangle_{\mu=2\tau}
+ \frac{1}{16\tau^2}
\langle \bar q\sigma \mbox{\rm g}G q\rangle_{\mu=2\tau}\,.
\label{FBHQL}
\end{equation}
 Using the sum rule in (\ref{FBHQL}) to eliminate the
coupling $\widehat f_B $ from (\ref{SRHQL}), we obtain
the invariant form factor for $B\rightarrow K^*\gamma$
\begin{equation}
    \widehat F_1(0) \simeq 5
\end{equation}
(with $\mu^2\sim 1 {\rm GeV}^2$),
which translates to the static limit of the decay form factor
(at the same $\mu^2$)
\begin{equation}
F_1(0) =
    \widehat F_1(0) /m_B^{3/2} \simeq 0.4\,.
\label{F1hql}
\end{equation}
This number appears to be 30\% higher than
the result in (\ref{ffactors}). The scale dependence reminds
of possible logarithmic corrections which we have ignored.
Note that the standard logarithmic factor corresponding to the
hybrid anomalous dimension of the interpolating current of the
$B$-meson cancels in the ratio of (\ref{SRHQL}) and (\ref{FBHQL}).
However, additional corrections can appear, similar to the
Sudakov-type corrections, which in the context
of inclusive radiative decays $B \to X_s + \gamma$ have been discussed
in \cite{ag1}. A related discussion of the perturbative corrections
produced by the hard gluon exchange can be found in \cite{Burdman}.
Since these corrections may exponentiate, the power  law
in (\ref{F1hql}) should be taken with caution.
A study of the perturbative two-loop corrections to the sum rule
would be welcome, but is beyond the tasks of this paper.

The contributions of higher twist to the sum rule (\ref{SR})
do not spoil the existence of the heavy-quark expansion.
In this case, again,
explicit enhancement factors $\sim m^2_b/t \sim O(m_b)$
are compensated by the smallness of the wave functions.
However, in contrast with the contributions of higher-dimension
condensates to
the sum rule in (\ref{FBHQL}), the
higher-twist contributions to (\ref{SRHQL}) generally
bring in factors $(\omega/\tau)$, which indicates that
the structure of the expansion is different.
We hope to return to this question in the future.

Thus, we conclude that the light-cone sum rules,
which we suggest in this paper, remain well-defined in the heavy
$b$-quark limit, since all the contributions appear to have
the same power behaviour at $m_b\rightarrow \infty$.
The same conclusion has been drawn in the work \cite{CZ-B}
for the case of semileptonic $B\rightarrow \pi e\nu$ decays.

%%%%%%%%%%%%%%%%%%%%%% SECTION 7 %%%%%%%%%%%%%%%%%%%%%%%%%%%%%%%%
\section{The $\bf q^2$-dependence and the interrelation of\newline
  radiative and semileptonic $\bf B$-decay form factors}

Up to this point, we have been discussing the radiative decay form factor
$F_1$ at $q^2=0$. Now we are in a position to calculate the $q^2$-dependence.
The interest in the $q^2$-dependence is due to the fact that
within most of the existing approaches the calculations are actually
done for  high values of $q^2$, close
to the maximum possible value $q^2_{\rm max} = (m_B-m_{K^*})^2$,
and then extrapolated to $q^2=0$ using plausible assumptions
like the pole-dominance approximation.
Within the QCD sum rule approach the $q^2$ dependence of
form factors can be calculated in a wide interval of $q^2$,
see e.g. \cite{BBD,BKR}.

The QCD sum rule for the form factor $F_1$ for finite values
of $q^2$ is obtained from  the one given in (\ref{SR}) by trivial
modifications. We obtain
\begin{eqnarray}
\lefteqn{\frac{f_B m_B^2}{m_b+m_q} 2 F_1(q^2)
e^{-(m_B^2-m_b^2)/t}  =}
\nonumber\\ &=&
 \int_0^1 du \frac{1}{u}\exp\left[-\frac{\bar u}{t}
\left(\frac{m_b^2-q^2}{u} + m_V^2\right)\right]
\theta\left[s_0-\frac{m_b^2-\bar u q^2}{u}-\bar u m_V^2\right]
\Bigg\{m_b f^\perp_V \phi_\perp(u)
\nonumber\\
&&\mbox{}+
 u m_V f_V g_\perp^{(v)}(u)
 +\frac{m_b^2+q^2-u^2m_V^2+ut}{ 4 u t} m_V f_V
 g_\perp^{(a)}(u) \Bigg\} \, .
\label{SRF1}
\end{eqnarray}
The region of applicability of this sum rule is restricted
by the requirement that the value of $q^2-m_b^2$ be sufficiently less
than zero, so that the correlation function (\ref{correlator})
is evaluated in the Euclidean region. In order not to introduce
an additional scale in the problem, we require that
$q^2-m_b^2\le (p+q)^2-m_b^2$, which translates to the condition
that $m_b^2-q^2$ is of order of the typical Borel parameter
$t\sim 5$--$8$ GeV$^2$. Thus, we end up with an upper bound
for the applicability of the QCD calculation
$q^2<15$--$17$ GeV$^2$, which is a few GeV$^2$ below the zero-recoil
point.
The results are shown in Fig. 6. It is seen that the stability
of the sum rules becomes worse with the increase of $q^2$.
The reason is that the omitted contributions of higher twists are
suppressed by factors $1/(m_b^2-u(p+q)^2)$ in the momentum
space, which translates into suppression by powers of $1/(ut)$
after the Borel transformation. At $q^2=0$ the average value of $u$
under the integral is of order 0.8, but it decreases to
$\sim 0.6$ at $q^2\simeq 15$ GeV$^2$. Thus, in order not to enhance
the higher twist effects one should somewhat
increase the value of the Borel parameter $t$ at high $q^2$.
This increase, however, potentially comes in conflict
with the requirement that the Borel parameter $t$ be small enough
to allow to separate the  $B$-meson contribution from the continuum.
Thus, in general, the light-cone sum rules for large values
of $q^2$ are less reliable than at $q^2=0$.
The calculation shown in Fig. 6 was done using the rescaled
Borel parameter  $t\rightarrow t/\langle u \rangle
 (q^2)$, where $\langle u \rangle (q^2)$ is
the average value of $u$ under the integral.
This allows for a certain improvement of the stability.

 For the radiative decays discussed here,
the sum rules do not contradict the pole-type behaviour
in the region $0 \le q^2 \le 10$ GeV$^2$:
\begin{equation}
     F_1(q^2) = \frac{F_1(0)}{1-q^2/m_{\rm pole}^2}
\label{poleformula}
\end{equation}
with the pole masses of order 3.8--4.2 GeV. A much better
quality of the fit is provided, however, by the dipole
formula
\begin{equation}
     F_1(q^2) = \frac{F_1(0)}{(1-q^2/m_{\rm dipole}^2)^2}
\end{equation}
with $m_{\rm dipole} = 5.6$, 5.4, 5.2, and 4.9 GeV for Fig.~6a, b, c,
and d, respectively.

In the case of $B\rightarrow K^*\gamma$ decays,
an important result was derived by
Isgur and Wise in \cite{IW} and by Burdman and Donoghue
 in \cite{Burdman}, where it has been shown that in the
heavy $b$-quark limit an exact relation exists between the form factor
$F_1$ and the semileptonic form factors defined by:
\begin{equation}
\langle K^*(p',\lambda)| \bar s\gamma_\mu(1-\gamma_5)b
|\bar B(p)\rangle  =
\frac{2V(q^2)}{m_B+m_{K^*}}\epsilon_{\mu\nu\rho\sigma}
e^{* (\lambda)\nu } p^\rho p^{'\sigma}
 -i(m_B+m_{K^*})A_1(q^2)e^{* (\lambda) }_\mu  +\ldots
\label{semileptonic}
\end{equation}
where the ellipses denote terms proportional to $(p+p')_\mu$ or
$q_\mu$. According to \cite{IW,Burdman}:
\begin{eqnarray}
     F_1^{B\rightarrow K^*\pg}(q^2) &=&
\frac{q^2+m_B^2-m_{K^*}^2}{2 m_B} \frac{V(q^2)}{m_B+m_{K^*}}
+\frac{m_B+m_{K^*}}{2 m_B} A_1(q^2)\,,
\label{IWrelation}
\end{eqnarray}
as far as the value of $q^2$ is sufficiently close to the
point $q^2_{\rm max} = (m_B-m_{K^*})^2 \simeq 19$ GeV$^2$.
Similar relations can of course be derived for other radiative
decays.
The central problem in using predictions based on the heavy quark
effective theory to the rare $B$-decays is whether this relation
is satisfied for small values of $q^2$, as suggested in
\cite{IW,Burdman}, and what is the size of the power $1/m_b$ corrections.
 Our main result in this section is that the relation in
(\ref{IWrelation}) is fulfilled in the QCD sum rule approach discussed
here to a good accuracy for
finite $b$-quark masses and for all momentum transfers.

The sum rules for the semileptonic form factors entering
(\ref{IWrelation}) can be derived in a similar way, repeating
the steps described in section~3. A simple calculation yields
\begin{eqnarray}
\lefteqn{\frac{f_B m_B^2}{m_b+m_q}
[m_B+m_V]A_1(q^2)
e^{-(m_B^2-m_b^2)/t}  = }
\nonumber\\&=&
 \int_0^1 du \frac{1}{u}\exp\left[-\frac{\bar u}{t}
\left(\frac{m_b^2-q^2}{u} + m_V^2\right)\right]
\theta\left[s_0-\frac{m_b^2-\bar u q^2}{u}-\bar u m_V^2\right]
\nonumber\\
&&\mbox{}\times
\Bigg\{\frac{m_b^2-q^2+u^2m_V^2}{2u}
 f^\perp_V \phi_\perp(u) + m_b f_V m_V g_\perp^{(v)}(u)
  \Bigg\} \,,
\nonumber\label{A1}\\
\lefteqn{\frac{f_B m_B^2}{m_b+m_q}
\frac{2V(q^2)}{m_B+m_V}
 e^{-(m_B^2-m_b^2)/t} =  }
\\&=&
\int_0^1 du \frac{1}{u}\exp\left[-\frac{\bar u}{t}
\left(\frac{m_b^2-q^2}{u} + m_V^2\right)\right]
\theta\left[s_0-\frac{m_b^2-\bar u q^2}{u}-\bar u m_V^2\right]
\nonumber\\
&&\mbox{}\times \Bigg\{f^\perp_V \phi_\perp(u)
+\frac{m_b  f_V m_V}{2ut}g_\perp^{(a)}(u) \Bigg\} \,.
\label{SRsemileptonic}
\end{eqnarray}
Note that the three sum rules in (\ref{SRF1}), (\ref{A1}), and
(\ref{SRsemileptonic}) involve the same wave functions as
specified in section~3.

It is now easy to check the Isgur-Wise relation by forming the ratio
\begin{equation}
   R(q^2)=F_1(q^2)/F_1^{\rm IW}(q^2)\,,
   \label{RIW}
\end{equation}
where $F_1(q^2)$ is the form factor of the radiative decay according to
(\ref{SRF1}), and $F_1^{\rm IW}(q^2)$ is the particular combination of
the semileptonic form factors, which enters on the r.h.s. of
(\ref{IWrelation}), and which is evaluated by using to the sum rules
(\ref{A1}) and (\ref{SRsemileptonic}).
The ratio (\ref{RIW}) is shown as a function of $q^2$ in Fig.~7 for
the two decays $B\rightarrow K^*\gamma$ and
$B\rightarrow \rho \gamma$, and is practically the same
for the two remaining decays $B_s\rightarrow \phi\gamma$ and
$B_s\rightarrow K^*\gamma$.
We find that the relation between radiative and semileptonic
form factors is satisfied with a high accuracy
for all  momentum transfers. For the physically relevant
point  $q^2=0$ a typical value of the ratio $R$ is about 0.92--0.95.
These predictions should be rather reliable,
 since many of the uncertainties in the individual decay rates are
greatly reduced in the ratios.

Since the semileptonic decay $B \to K^* + \ell +\nu_\ell $ does not
occur in nature, tests of the Isgur-Wise relation must take into
account the correction for the $SU(3)$ breaking between $B \to K^*$ and
$B \to \rho$ semileptonic form factors. From our sum rule
calculation we get
\begin{eqnarray}
     A_1(0)^{B\to\rho}/A_1(0)^{B\to K^*} &=& 0.76 \pm 0.05 \,,
\nonumber\\
     V(0)^{B\to\rho}/V(0)^{B\to K^*} &=& 0.73 \pm 0.05 \,.
\end{eqnarray}
As a by-product of our analysis, we have derived the $q^2$ dependence
of the semileptonic decay form factors, $V$ and $A_1$, which
is shown in Fig.~8 for the transition $B \to \rho + \ell +\nu_\ell$.
In Table~1 the values for $V(q^2=0)$ and $A_1(q^2=0)$ are
compared with the corresponding results from
other approaches \cite{Semi1}--\cite{Semi4}. We find a satisfactory
agreement with the quark models, while our results are somewhat
lower than previous QCD sum rule estimates, which were done
using the traditional approach. In agreement with \cite{Semi1}
we obtain a steeper increase of the form factor $V(q^2)$
with $q^2$, compared to $A_1(q^2)$, although values of
the slopes come out to be different.

 Predictions of the traditional  sum rules should be
more reliable at the values
$m_b^2-q^2= O(m_b)$  than at $q^2=0$, since in this region
they do not suffer from large contributions
of the operators of high dimension, see the detailed discussion in the
appendix.
In the light-cone sum rules the situation is opposite, so that
our results are to a large extent complementary to
the calculation in \cite{Semi1,Semi2}. It is worth while to note
that the difference in predictions
is minimal at intermediate $q^2\sim 5$--$10$ GeV$^2$, i.e. exactly
in the region where both the approaches should work well.
\begin{table}[h]
\begin{center}
\caption{Form factors for the semileptonic decay $B \to \rho + \ell +
\nu_\ell$}
\vspace{0.6cm}
\begin{tabular}{|c|c|c|}
\hline
Reference  & $A_1(0)^{B \to \rho}$ & $V(0)^{B \to \rho }$  \\
\hline
This paper  & $0.24\pm 0.04$  & $0.28 \pm 0.06$ \\
\protect\cite{Semi1}$^{a}$ & $0.5 \pm 0.1$  & $0.6 \pm 0.2$ \\
\protect\cite{Semi1A}$^{a}$ & $0.35 \pm 0.16$  & $0.47 \pm 0.14$ \\
\protect\cite{Semi2}$^{b}$ & $0.28$  & $0.33$ \\
\protect\cite{Semi3}$^{b}$ & $0.05$  & $0.27$ \\
\protect\cite{Semi4}$^{c}$ & $0.21$  & $1.04$ \\
\hline
\end{tabular} \\
\vskip0.3cm
\parbox[t]{6.5cm}{ \small{
a: QCD sum rules\\
b: Quark models\\
c: HQET + chiral perturbation theory\\ }}
\end{center}
\end{table}

%%%%%%%%%%%%%%%%%%%%%% SECTION 8 %%%%%%%%%%%%%%%%%%%%%%%%%%%%%%%%

\section{Summary and Conclusions}
Motivated by the discovery of the electromagnetic penguins through the
decay $B \to K^* + \gamma$, we have presented an alternative method of
calculating the transition form factors in this and related decays
using the approach of QCD sum rules on the light cone.  Our
calculations give $F_1^{B \to K^* \pg}=0.32 \pm 0.05$;
combined with the estimates of the inclusive branching ratio $BR(B \to
X_s + \gamma)=(3.0\pm1.2) \times 10^{-4}$ in the Standard Model
\cite{ag5}, this yields $BR(B \to K^* + \gamma)=(4.8\pm 1.5)\times
10^{-5}$, which is in reasonably good agreement with the observed
branching ratio $(4.5\pm 1.5\pm 0.9)\times 10^{-5}$ measured by the
CLEO collaboration.

\par
We use the same method to calculate the
related form factors in the decays $B_{u,d} \to \rho + \gamma$,
$B_d \to \omega + \gamma, ~B_s \to \phi + \gamma$ and $B_s \to K^* +
\gamma$. These decays can be related to the already discussed and
measured decay $B_{u,d} \to K^* + \gamma$. To this end, we have evaluated
in detail the numerical effect of the $SU(3)$-breaking in the
form factors.
This fixes the decay width  $\Gamma (B_s \to \phi + \gamma)$,
which involves the same CKM matrix element $\Vtsabs$ as the decay
width $\Gamma(B_{u,d} \to K^* + \gamma)$. Since the form factors
for the two decay modes are comparable in size, see Fig.~3 and
eq.~(\ref{ffactors}), and since the lifetimes of the
$B_u,B_d$ and $B_s$ mesons are expected to be very similar --- an
expectation which is consistent with present experiments --- we estimate
that $BR(B_s \to \phi + \gamma) \simeq BR(B_{u,d} \to K^* + \gamma)$.

\par
More interesting, however, are the CKM-suppressed decay modes
discussed above. A measurement of both the CKM-allowed and
CKM-suppressed radiative decays $B \to V + \gamma$ would give direct
information on the CKM matrix element ratio $\Vtdabs/\Vtsabs$. The
form factors needed for such determination are plotted in Fig. 4,
showing the ratios of $F_1^{B \to \rho \pg}/F_1^{B \to K^* \pg}$,
$F_1^{B_s \to K^* \pg}/F_1^{B_s \to \phi\pg}$
and $F_1^{B_s \to K^* \pg}/F_1^{B_{u,d} \to K^* \pg}$. The
numerical results for the form factor ratios are given in
eqs.~(\ref{ratios})--(\ref{ratioks}). The ratios of the branching
ratios $BR(B \to \omega + \gamma)/BR(B \to K^* + \gamma)$ and $BR(B_s
\to K^* + \gamma)/ BR(B \to K^* + \gamma)$ as functions of
$\Vtdabs/\Vtsabs$ are shown in Fig. 9 ($B=B_u$ or $B_d$.).  As already
stressed, these ratios are practically independent of $m_t$ and of the
QCD scale parameter $\mu$.  The dependence of the ratios on the
various parameters endemic to the QCD sum rule approach also becomes
rather mild, as opposed to the predictions for the partial decay
widths themselves.  Therefore, the two curves shown in Fig.~9 should
give a fair representation of the residual theoretical error, and they
allow a rather clean determination of the ratio of the CKM matrix elements
$\Vtdabs/\Vtsabs$ as and when the required data for the ratios of
branching ratios become available.

\par
 It should be pointed out here that similar
relations have been derived for the ratios of the inclusive branching
ratios $BR(B \to X_d + \gamma)/BR(B \to X_s + \gamma)$ in \cite{ag3}, and
for the ratios of the $B^0$--$\overline{B^0}$
 mixing parameters $x_d$ and $x_s$
\cite{AL92}. In particular, the Standard Model predicts:
\begin{equation}
\frac{x_d}{x_s} = \frac{\tau_{B_d}M_{B_d}\left(\fbb\right)}
{\tau_{B_s}M_{B_s}\left(\fbbs\right)}
\left\vert \frac{V_{td}}{V_{ts}} \right\vert^2, \label{xratio}
\end{equation}
where $B_{B_d}$ and $B_{B_s}$ are the hadronic bag constants. Since
all dependence on the $t$-quark mass and the QCD renormalization of the
box amplitudes drops out, we are left with the
square of the ratio of CKM matrix elements, multiplied by a factor that
reflects $SU(3)_{\rm flavour}$-breaking effects.
With the known values of $x_d$ and $BR(B \to K^* + \gamma)$,
a measurement of $x_s$ and of the  CKM-suppressed radiative decays
would provide valuable quantitative tests on the CKM matrix and,
in particular, its unitarity.

\par
In order to obtain an estimate of the absolute branching ratios for
the CKM-suppressed decays, we use the constraints on the matrix element
$\Vtdabs$ that already exist from the measurement of the mixing ratio $x_d$.
The present value $x_d = 0.71 \pm 0.07$ \cite{Danilov} yields the
model dependent bound $0.005 < \Vtdabs < 0.012$ \cite{Danilov}.
This in turn gives
\begin{equation}
  0.10 \leq \frac{\Vtdabs}{\Vtsabs} \leq 0.33 ,
\end{equation}
where we have used the CKM
 unitarity constraint and recent data giving $\Vtsabs \simeq \Vcbabs
=0.042 \pm 0.005$ \cite{Cassel}. We can combine this bound with Fig.~9
to predict:
\begin{equation}
 0.01 \leq \frac{BR(B \to \omega + \gamma )}{BR(B \to K^* + \gamma )}
  \leq 0.10,
\end{equation}
with the ratio
$BR(B \to \rho + \gamma)/BR(B \to K^* + \gamma)$ bounded likewise.
While this gives a ball-park estimate for the branching
ratios in question, it is clear from this numerical exercise that a
measurement of the above ratio within a factor 2 would enormously
reduce the present (model-dependent) uncertainty on the CKM matrix
element ratio $\Vtdabs/\Vtsabs$.

The main theoretical content of this paper is in pointing out and
providing general arguments for the light-cone QCD sum rule approach
to be a more adequate tool for the study of heavy meson decay form
factors compared to the traditional approach. The principal input
in the light-cone sum rules are the wave functions of vector mesons,
and the accuracy of our calculation can be substantially improved
by taking into account the wave functions of twist 3 and twist 4.
For the case of the $\pi$-meson, a complete set of wave functions
to this accuracy is given in \cite{BF2}. For vector mesons,
this problem is not solved yet, and is of
acute practical interest. Some new
results concerning the interrelation of the transverse and
longitudinal wave functions of vector mesons are given in section 4.

\par
We have also studied the heavy-quark mass dependence of $F_1$
 in the region $m_c \leq m_Q \leq m_b$.
This is useful for comparison with lattice-QCD results where the
heavy quark systems at present  are calculated in the vicinity of $m_c$ and
extrapolated to $m_b$. We find that the transition form factor $F_1$
scales approximately in this region as $F_1 \sim 1/m_b$, and has a
value $F_1 = 0.85 \pm 0.13$ for $m_B$ equal to the $D$-meson mass, i.e.
for the (unphysical) process $D \to K^* + \gamma$.

\par
Apart from the phenomenologically interesting value of the form
factor $F_1^{B \to V \pg}$ at the on-shell point $q^2=0$,
we have investigated the behaviour of $F_1$ as a function of $q^2$.
The case $q^2 \not=0$ contributes to the decays $B \to V + \ell^+ \ell^-$
via off-shell photons. In addition,
we have derived the light-cone QCD sum rules for the form factors
$A_1(q^2)$ and $V(q^2)$, describing the semileptonic decay $B \to
\rho +\ell\, \nu_\ell$.
It will be interesting to compare them with data on
the semileptonic decay $B \to \rho +\ell\,\nu_\ell$, as and when these
become available. Another use of this sum rules is to check the
relation between the radiative decay form factor
 $F_1^{B \to K^* \pg}$ and the semileptonic form factors in
 $B \to \rho +\ell\, \nu_\ell$, which was derived in \cite{IW, Burdman}
using the
heavy-quark symmetry and for $q^2$ near the kinematic point
$q^2_{\rm max}$. We have shown that the ratio $F_1(q^2)/F^{IW}_1(q^2)$
evaluated in the QCD sum rule approach remains very close to unity
in the complete kinematic range of $q^2$ which is available in the
CKM-suppressed semileptonic decays.

\par
In summary, we have given a number of predictions here in the crucial
area of exclusive radiative and semileptonic $B$-decays, involving the
CKM-suppressed and CKM-allowed transitions using QCD sum rules on the
light-cone. These predictions are specific to this approach and it will
be instructive to confront them with data. The existing data on
$BR(B \to K^* + \gamma)$ are reasonably well explained, and we argue that
our results allow to quantify the ratios of CKM-allowed and CKM-suppressed
exclusive radiative $B$-decays. This would prove useful in determining the
CKM ratio $\Vtdabs/\Vtsabs$ in a fairly model-independent way.

\par
{\bf Acknowledgements}\\
  We acknowledge helpful discussions with David Cassel, Michael Danilov,
Thomas Mannel, Stephan Narison and Matthias Neubert. One of us (V.M.B.)
would like to thank John Ellis and the CERN TH division for their
hospitality.
\newpage
\begin{flushright}
{\Large \bf Appendix}
\end{flushright}
\bigskip
\section*{Three-point sum rule }
\renewcommand{\theequation}{A.\arabic{equation}}
\setcounter{equation}{0}

In this appendix we compare our approach,
for the particular case of $B\rightarrow K^*\gamma$ decays,
with the results of the standard QCD sum rule analysis
\cite{Aliev,Dom93,Ball93}.
Here, the starting point is the operator product expansion
for the three-point function involving the penguin operator
and the interpolating currents for both the $B$-meson and the $K^*$-meson
\begin{eqnarray}
i^2\int dx\,dy\, e^{ipx+iqy}
\langle 0 | T\{ [\bar d \gamma_\alpha s](x)
[\bar s \sigma_{\mu\nu}q_\nu b](y)
[\bar b i \gamma_5 d](0)\}|0\rangle
\label{3pt}
\end{eqnarray}
in which the contribution of interest is identified with
the one having poles at $p^2 = m_{K^*}^2$ and $(p+q)^2 = m_B^2$.
The standard treatment, details of which we are not going to present
here, leads to the following sum rule:
\begin{eqnarray}
\lefteqn{
\frac{f_{K^*}m_{K^*}f_B m_B^2}{m_b}\,2 F_1^{B\rightarrow K^*\pg}(0)
e^{-m_{K^*}^2/t_1 } e^{-(m^2_B-m^2_b)/t_2} =  }
\nonumber\\
&=& \frac{3}{4\pi^2}
\int_0^{s_0^K} ds_1\, e^{-s_1/t_1}
\int_{m_b^2}^{s_0^B} ds_2 e^{-(s_2-m^2_b)/t_2}
\theta(s_2-s_1-m^2_b) \rho(s_1,s_2) -
(m_s+m_b)\langle \bar d d\rangle
\nonumber\\
&&\mbox{} +\frac{1}{12} \langle \bar d\sigma g G d\rangle
\left\{
\frac{3m_b^2(m_s+m_b)}{t_2^2} + \frac{2m_b^2(2m_b+m_s)}{t_1t_2}
+\frac{6 m_b+8m_s}{t_2} -\frac{2m_s}{t_1}
\right\},
\label{SR3}
\end{eqnarray}
where $\rho(s_1,s_2)$ is the spectral density
of the triangle diagram in Fig. 10a
\begin{equation}
\rho(s_1,s_2) = \frac{m_b^4 s_1}{(s_2-s_1)^3}
+m_b m_s \frac{s_2-s_1-m_b^2}{(s_2-s_1)^2} \,,
\end{equation}
and we have calculated also the contribution of the quark
condensate in Fig. 10b, and of the mixed condensate in
Fig. 10c (a typical graph is shown).
The contributions of the gluon condensate and of the four-quark
condensate are negligible \cite{Ball93}. In contrast with ref.
\cite{Dom93}, we give the answer with $O(m_s)$ corrections
included. To the accuracy that these corrections are put
to zero, our sum rule  agrees with the one given in \cite{Dom93},
but for the sign in front of the third term in the contribution
of the mixed condensate. Numerically, the difference is
negligible. Note that
in contrast to the light-cone sum rules  we have now two Borel
parameters, $t_1$ for the $K^*$-meson and $t_2$ for the
$B$-meson, which should be taken two times larger than the
Borel parameters in the corresponding two-point sum rules
(cf. \cite{BBD}).
The quantity
  $s_0^{K}=1.7$  GeV$^2$ is the continuum threshold in the sum
rules for the $K^*$-meson, and $m_s$ is the mass of the strange
quark, which is of order  $m_s = 150$ MeV. The other entries
have been specified in the text.

A numerical treatment of the sum rule in (\ref{SR3}) yields
values of the form factor of order
\begin{equation}
  F_1^{B\rightarrow K^*\pg}(0)
 \simeq 0.5\mbox{--}0.6
\label{answer-SR3}
\end{equation}
which is in agreement with the analysis in \cite{Aliev,Ball93}.
The lower value $\sim 0.38$ quoted in \cite{Ball93} was obtained by
the rescaling of a similar value $\sim$ 0.55 from the scale of
 the typical Borel parameter to the
scale $\mu=m_b$, using the large anomalous dimension of the
penguin operator (\ref{O7}). We have not understood the reasoning
for such a rescaling, since the renormalization group treatment
of the effective Hamiltonian in (\ref{Hamiltonian}) makes sense
at scales $\mu>m_b$ only. A somewhat smaller value obtained in
ref. \cite{Dom93} is due to a smaller value of the quark condensate
and due to a higher value $f_B=180$ MeV used in \cite{Dom93}.

The  value in (\ref{answer-SR3})
appears to be considerably larger than our result in
(\ref{ffactors}). We are going to claim, however, that
 the sum rule in (\ref{SR3}) is
ill-behaved in the limit of heavy $b$-quark, and is less reliable.

An inspection of the sum rule in (\ref{SR3}) shows that various
terms in it have a different behaviour
in the limit  $m_b\rightarrow\infty$.
One easily finds that the perturbation theory contribution to
(\ref{SR3}) yields the form factor $F_1(0)\sim O(m_b^{-3/2})$,
the quark condensate produces a contribution of order
$m_b^{1/2}$, and the mixed condensate contribution is of order
$m_b^{3/2}$. Thus, the sum rule (\ref{SR3}) blows up in the limit
$m_b\rightarrow\infty$.

This problem has actually been known for a long time, and was
originally found in the QCD sum rule calculations
of the pion form factor \cite{IS,NR}.
In technical terms, the problem appears because the operator
product expansion for three-point functions involves the
dimensionless parameter $(m^2-q^2)/t$, increasing powers of
which multiply the contributions of local operators of
higher dimensions. In the case of light quark systems, $m^2=0$
and the Borel parameter is of order 1 GeV$^2$. Thus the expansion
breaks down at sufficiently large values of $q^2$.
In the case of heavy--light mesons $t\sim O(m_b)$, and for $q^2=0$
the operator product expansion involves increasing powers of the
heavy-quark mass.  In the present case, we observe
a considerable enhancement of the contribution of the mixed
condensate, and conclude that higher-order
corrections, e.g. proportional to the dimension-7 condensate
$\langle\bar \psi g^2 G^2 \psi\rangle$, can influence the
result significantly. It has been claimed recently
\cite{qGGq} that this condensate is much larger than its
factorized value.

In physical terms, the reason of the difficulty with
the traditional three-point sum rules in the $m_b\rightarrow\infty$
limit is that
the operator product expansion does not take into account
the effect of the finite correlation length between the quarks
in the physical vacuum. A possible remedy, suggested in \cite{MR},
is to use the expansion in non-local condensates, which take
into account this effect in a model-dependent way, see e.g.
\cite{Rad}.
Following Radyushkin, we write down the non-local quark
condensate in the Borel representation:
 \begin{equation}
\langle \bar d(x) d(0)\rangle =
\langle \bar d d\rangle \int_0^\infty d\nu\, e^{\nu x^2/4}
 f(\nu).
\label{nlc}
\end{equation}
Moments of the function $f(\nu)$ are determined by vacuum expectation
values of local operators
\begin{eqnarray}
\int_0^\infty d\nu\,f(\nu) &=&1\,,
\nonumber\\
\int_0^\infty d\nu\,\nu f(\nu) &=& \frac{1}{4}
\langle \bar d \sigma \mbox{\rm g}G d\rangle/\langle \bar d d\rangle
\simeq 0.2 ~\mbox{GeV}^2\,,
\end{eqnarray}
and similar relations for higher moments.
An effect in the sum rule (\ref{SR3})
is that in the term proportional to
the quark condensate one should make a replacement
\begin{equation}
\langle \bar d d\rangle \rightarrow \langle \bar d d\rangle
\int_0^t d\nu\,f(\nu)\exp\left
\{-\frac{m_b^2}{t_2}\frac{\nu}{t-\nu}\right\}
\label{nlc2}
\end{equation}
where $t\equiv t_1t_2/(t_1+t_2)$, and discard the contribution
of the mixed quark--gluon condensate, since
its contribution is partly included in (\ref{nlc}).
In the limit of the heavy $b$-quark mass the integral over $\nu$
in (\ref{nlc2}) is dominated by contributions of small $\nu$;
thus the behaviour of $f(\nu)$ at $\nu\rightarrow 0$ is of
crucial importance. Note that no information about this
behaviour can be extracted from known values of a few first moments.
In a confining theory such as QCD one should expect that
 correlation functions
 decrease exponentially at large distances in Euclidean space,
\begin{equation}
\langle \bar d(x) d(0)\rangle \stackrel{x^2\rightarrow -\infty}{\sim}
\exp\{-a\sqrt{-x^2}\}\,,
\end{equation}
where $a$ is the correlation length. Existence of a finite correlation
length in the physical vacuum is a part of a common wisdom about
confinement and a starting point of many
existing models of the vacuum (see e.g. \cite{Dosch}).
It is easy to show that the finite correlation length implies
quite generally
the exponential behaviour of the function $f(\nu)$ at small $\nu$:
\begin{equation}
    f(\nu) \sim e^{-a^2/\nu}.
\end{equation}
Inserting this behaviour in
(\ref{nlc2}) and taking the limit $m_b\rightarrow\infty$
we obtain to the exponential accuracy the contribution of the
quark condensate to be
\begin{equation}
        \sim \exp\{-2am_b/\sqrt{t_1t_2}\} \,.
\end{equation}
Since $t_1 = O(1)$ and $t_2 =O(m_b)$, this result means that
the contribution of the quark condensate decreases as
$\exp\{-\sqrt{m_b}\}$. Hence, the perturbative contribution
dominates the sum rule in this limit, yielding the same
behaviour $F_1(0)\sim m_b^{-3/2}$ which is intrinsic to the
light-cone sum rules in the text.
Note that this result is a quite general  consequence of the
existence of a finite correlation length in the QCD vacuum.
It is actually not surprising, since the power behaviour of
exclusive process with large momentum transfers
\cite{exclusive,BLreport} is given by perturbative diagrams.
In the present case, the diagrams with hard gluon exchange yield
the same power behaviour as the ones considered here, see \cite{CZ-B}.

To summarize,
 we argue that the light-cone sum rule written in (\ref{SR})
remains well-defined
in the limit of the heavy $b$-quark mass, similar to the
 traditional QCD sum rules for the two-point functions,
which can be written directly in the heavy-quark effective theory
\cite{BBBD}. The heavy-quark limit does not exist, however,
in the standard sum rules for three-point
functions, as the one in (\ref{SR3}). This
property of the light-cone sum rules argues in favour of
our approach.

%%%%%%%%%%%%%%%%%%%%% REFERENCES %%%%%%%%%%%%%%%%%%%%%%%%%%%%%%%%

\newpage
\section*{Figure captions}
\begin{description}

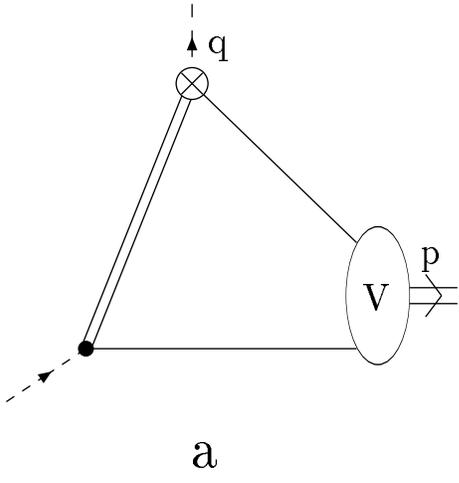
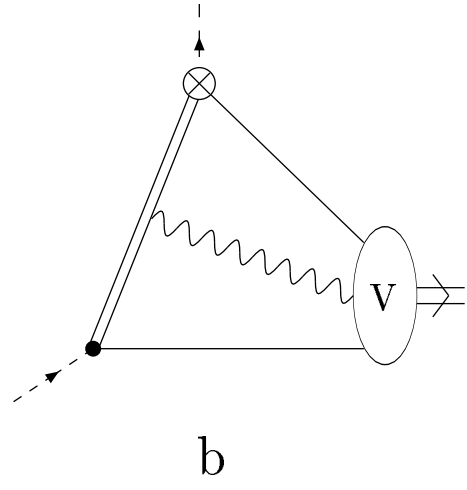
\item [Fig. 1] The leading contribution (a) and the gluon correction
            (b) to the correlation function in (\ref{correlator}).

\item [Fig. 2] The leading-twist wave function $\phi_\perp$
(\ref{twist2}) from the sum rule analysis in \cite{CZreport}
at the scale $\mu^2 =5$ GeV$^2$ for (a) $\rho$- (solid line) and $\phi$-
(dashed dotted line) mesons and (b) for the $K^*$-meson.
The wave function for $K^*$ is also
shown at the low scale  $\mu^2 =1$ GeV$^2$ (dashed line).
The dotted lines correspond to the asymptotic wave function
$\phi_\perp^{\rm as} = 6 u(1-u) $.

\item [Fig. 3] Stability plots for the sum rule in (\ref{SR})
as a function of the Borel parameter for $B\rightarrow K^*\gamma$,
$B\rightarrow \rho\gamma$, $B_s\rightarrow \phi\gamma$, and
$B_s\rightarrow K^*\gamma$, see  a, b, c and d, respectively.
Dashed lines correspond to $s^B_0=35\,~\mbox{GeV}^2$ and dotted lines
to $s^B_0=33\,~\mbox{GeV}^2$ (with $s_0^{B_s}$ given in (\ref{s0Bs})).
For each value of $s_0$ curves are shown
for $m_b=4.6 \,~\mbox{GeV}$ (upper curve),
$m_b=4.7\,~\mbox{GeV}$ (middle) and $m_b=4.8\,~\mbox{GeV}$
 (lowest curve).

\item [Fig. 4] Ratios of the form factors $F_1(0)$ for the
processes (a) $B\rightarrow \rho\gamma$ and $B\rightarrow K^*\gamma$,
(b) $B_s\rightarrow K^*\gamma$ and $B_s\rightarrow \phi\gamma$, and
(c) $B_s\rightarrow K^*\gamma$ and $B\rightarrow K^*\gamma$.
Dashed lines correspond to $s^B_0=35\,~\mbox{GeV}^2$ and dotted lines
to $s^B_0=33\,~\mbox{GeV}^2$ (with $s_0^{B_s}$ given in (\ref{s0Bs})).
The corresponding upper curves are for $m_b=4.6 \,~\mbox{GeV}$ and the
lower ones for $m_b=4.8\, ~\mbox{GeV}$.

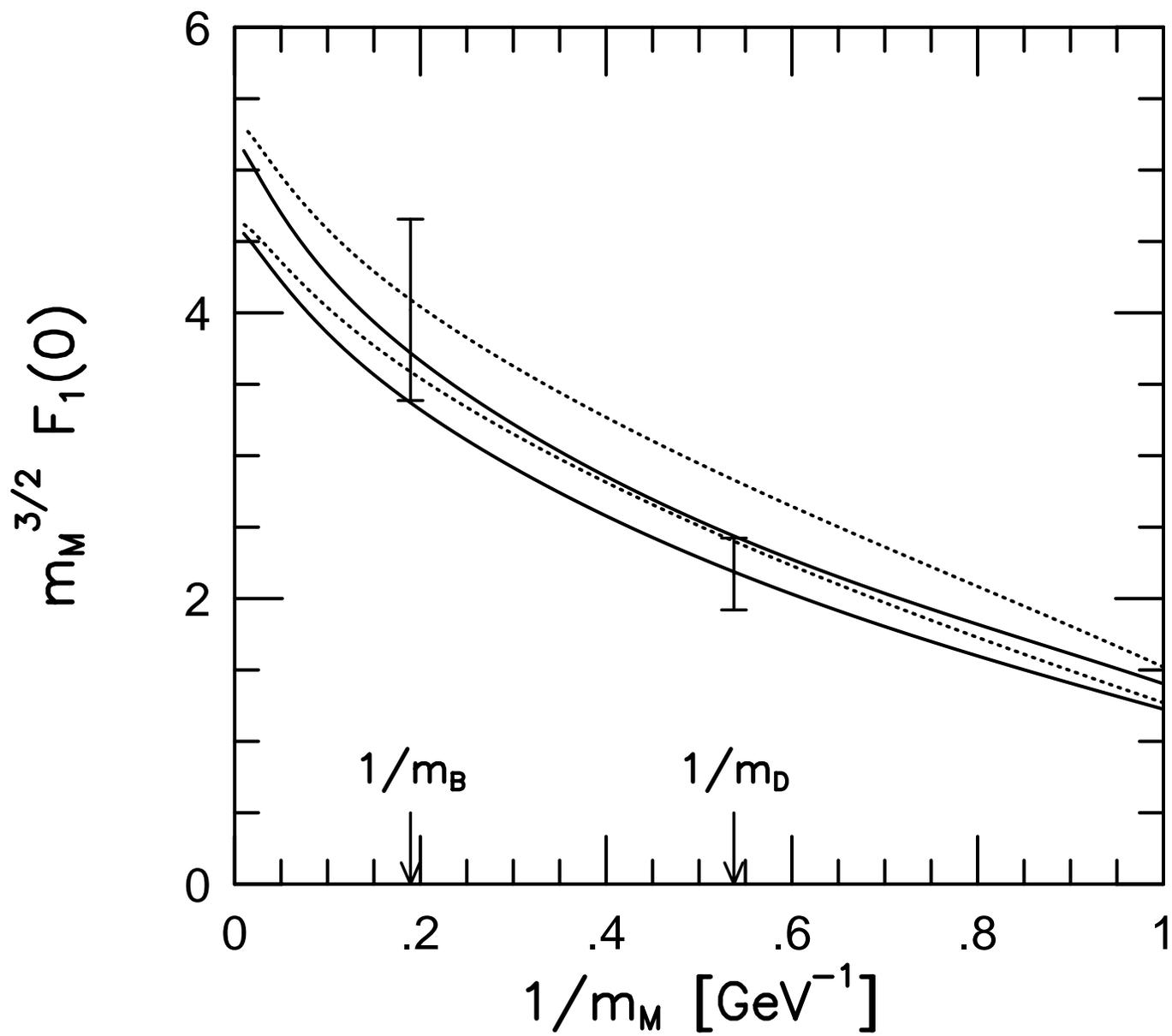
\item [Fig. 5] Dependence of the form factor $F_1(0)$ on the
 quark mass.
The various parameters are scaled according to (\ref{hqetscaling}).
Solid lines correspond to $\tau = 1.0\,~\mbox{GeV}$ and the dotted
lines to $\tau = 0.6$ GeV. The corresponding upper curves are
for $\bar{\Lambda} = 600\,\mbox{MeV}$ with $\omega_0 = 1.2\, \mbox{GeV}$,
and the lower ones for $\bar{\Lambda} = 500\,\mbox{MeV}$
with $\omega_0= 1.0\,\mbox{GeV}$.
For comparison the results corresponding
to the parameters discussed in section 5 are indicated by vertical bars.

\item [Fig. 6] Momentum dependence of the form factors $F_1(q^2)$ for
(a) $B\rightarrow K^*\gamma$, (b) $B\rightarrow \rho\gamma$, (c)
$B_s\rightarrow \phi\gamma$, and (d) $B_s\rightarrow K^*\gamma$.
Dotted lines correspond to $t = 5\,\mbox{GeV}^2 /\langle u\rangle$ and
dashed lines to $t = 8\,\mbox{GeV}^2 /\langle u\rangle$.
The corresponding upper curves are for $m_b=4.6\, ~\mbox{GeV}$ and the
lower ones for $m_b=4.8 \,~\mbox{GeV}$.

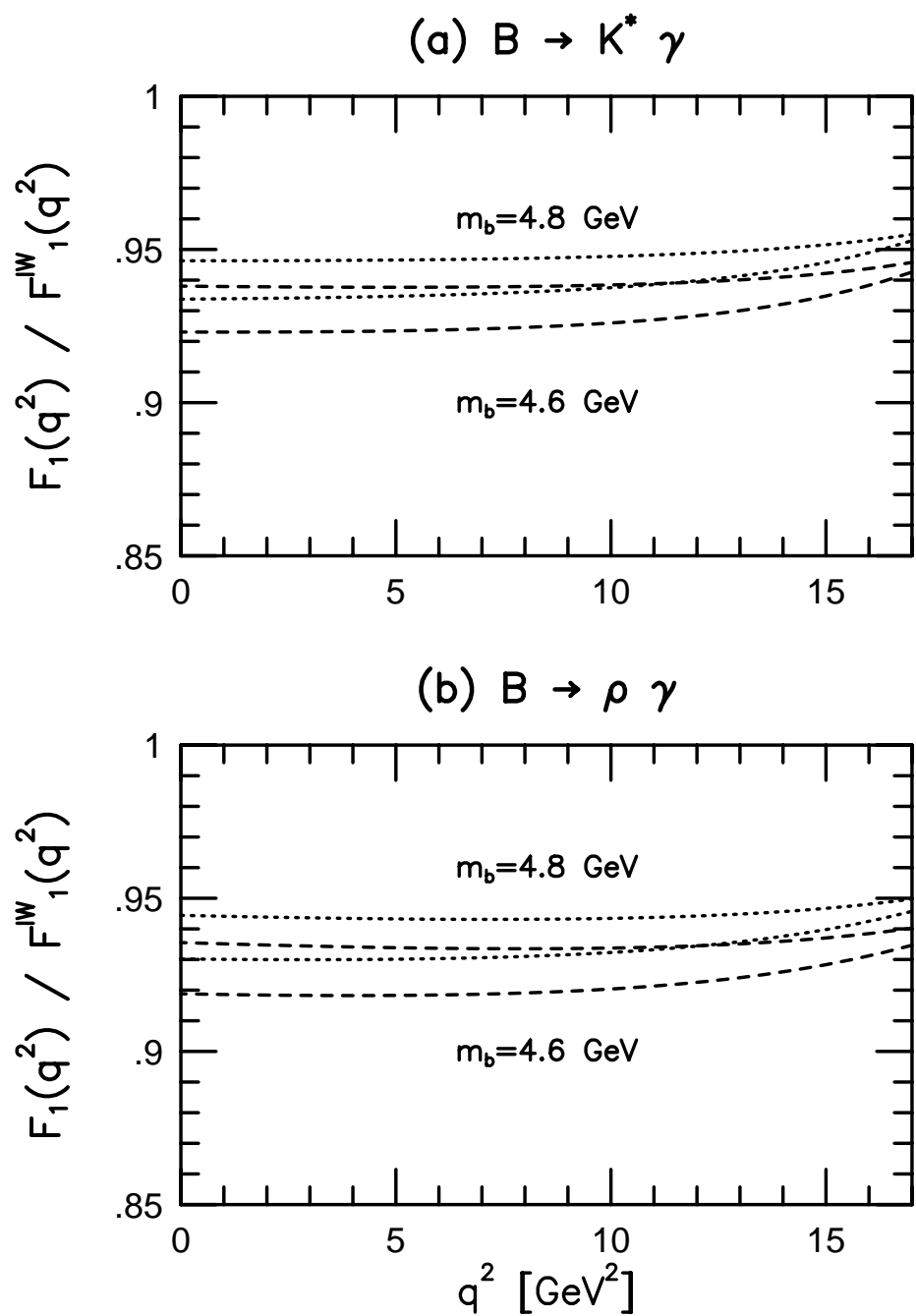
\item [Fig. 7] Momentum dependence of the ratio of radiative
and semileptonic form factors
(a) $B\rightarrow K^*\gamma$ and (b) $B\rightarrow \rho\gamma$.
Dotted lines correspond to $t = 5\, ~\mbox{GeV}^2 /\langle u\rangle$ and
dashed lines to $t = 8\, ~\mbox{GeV}^2 /\langle u\rangle$.
The corresponding upper curves are for $m_b=4.8\,~\mbox{GeV}$ and the
lower ones for $m_b=4.6\,~\mbox{GeV}$.

\item [Fig. 8] Momentum dependence of the form factors $A_1(q^2)$ (a)
and $V(q^2)$ (b) for $B\rightarrow \rho\gamma$.
Dotted lines correspond to $t = 5 ~\mbox{GeV}^2 /\langle u\rangle$ and
dashed lines to $t = 8 \,~\mbox{GeV}^2 /\langle u\rangle$.
The corresponding upper curves are for $m_b=4.6\, ~\mbox{GeV}$ and the
lower ones for $m_b=4.8\, ~\mbox{GeV}$.

\item [Fig. 9] Ratio of the branching ratios
 $BR(B\to \omega\gamma)/BR(B\to K^*\gamma)$, where $B=B_u$ or $B_d$,
and $BR(B_s\to K^* \gamma)/BR(B \to K^*\gamma)$,
    as a function of the CKM matrix elements
$\vert V_{td}\vert
/ \vert V_{ts}\vert$. The two lines correspond to the errors given in
(\ref{ratios}).

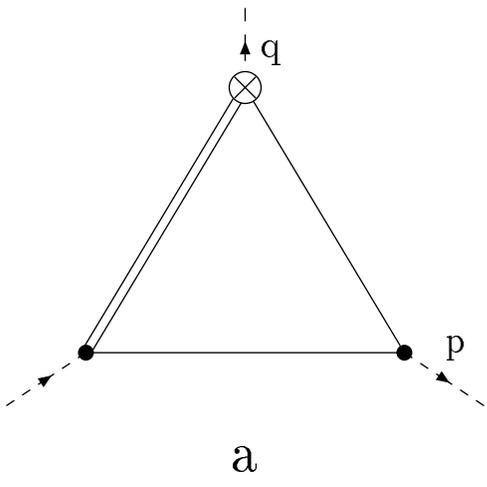
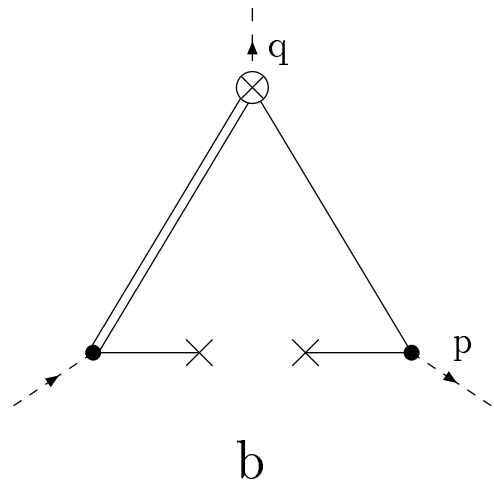
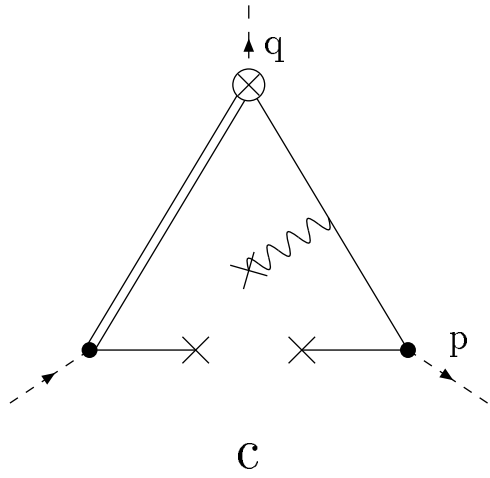
\item [Fig. 10] Contributions of perturbation theory and of
  vacuum condensates to the sum rule in (\ref{SR3}).

%\item 
Fig. 11Y Stability plot for the sum rule in (\ref{SR3})
%for three values of $m_b$. The other parameters
%are specified in the text.
\end{description}
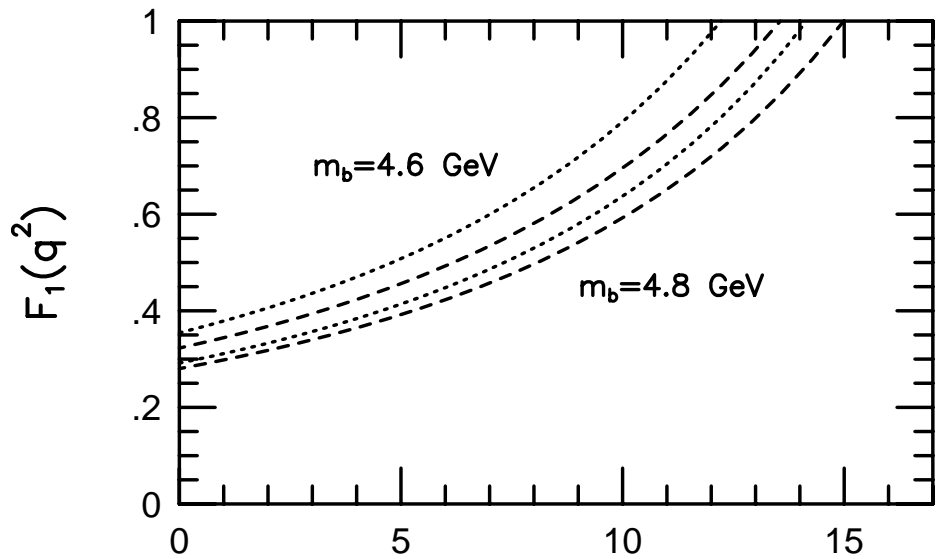
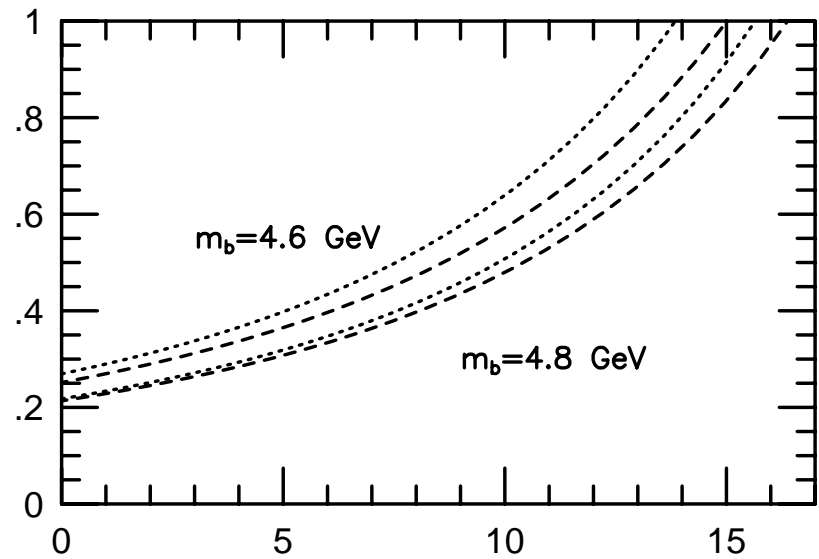
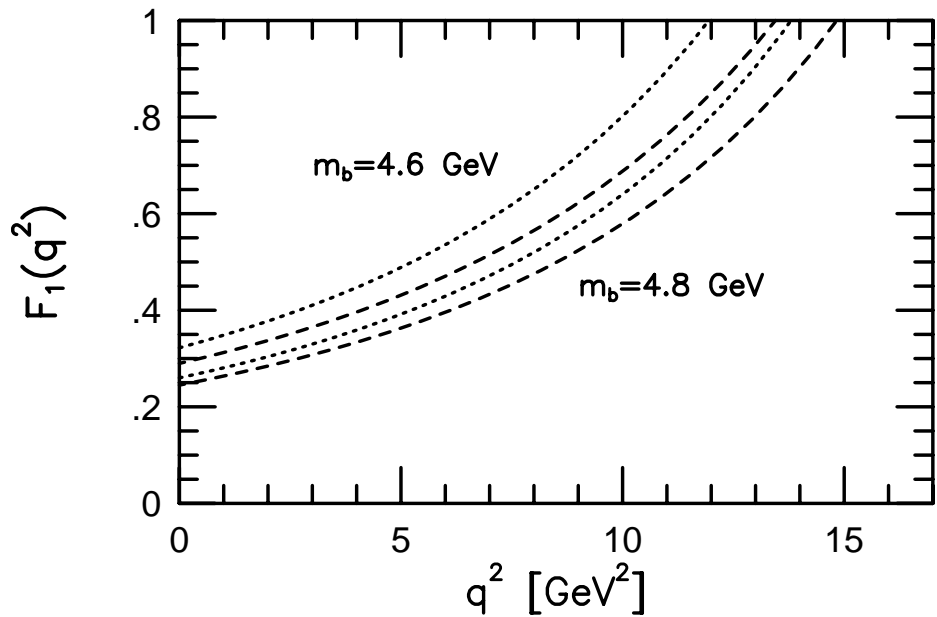
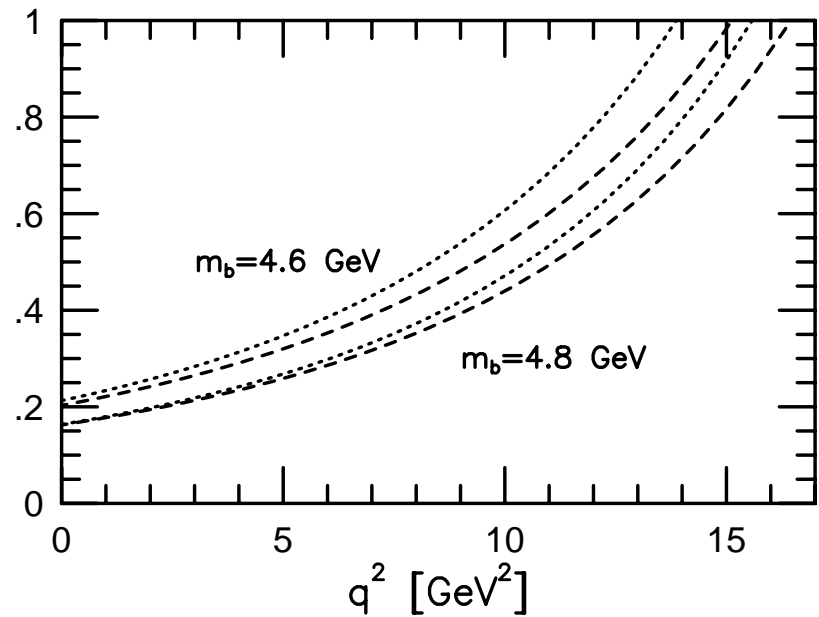
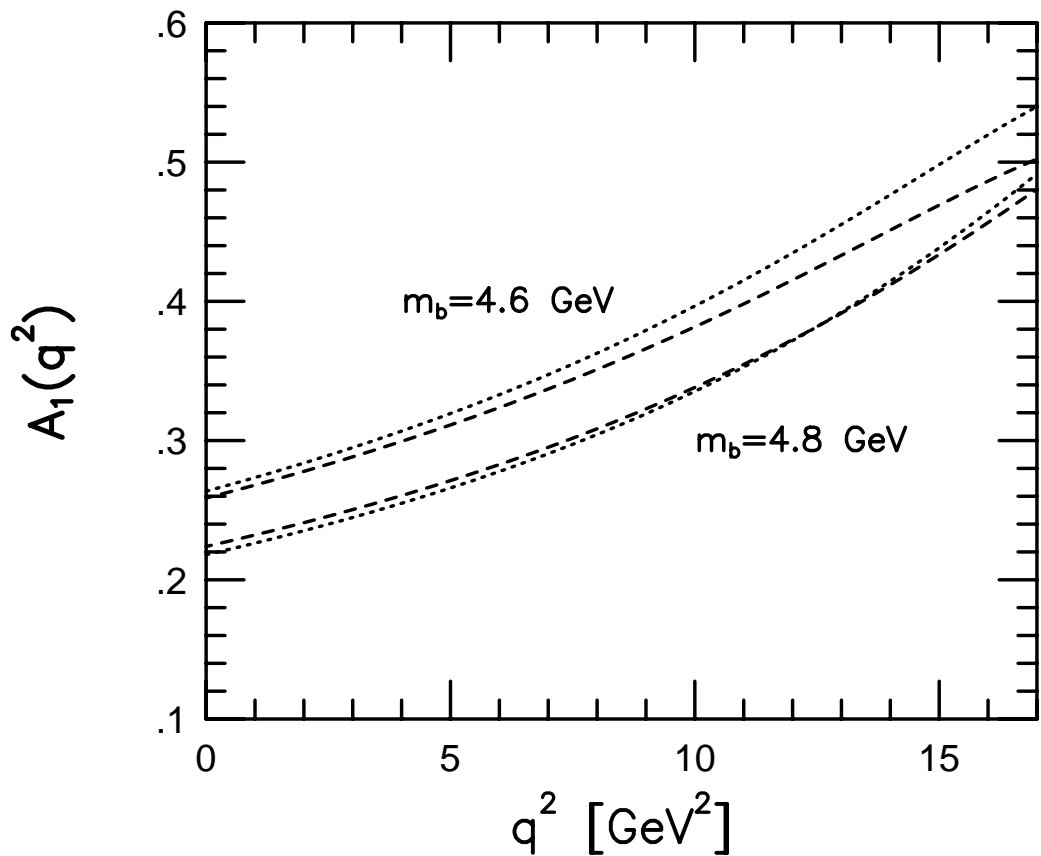
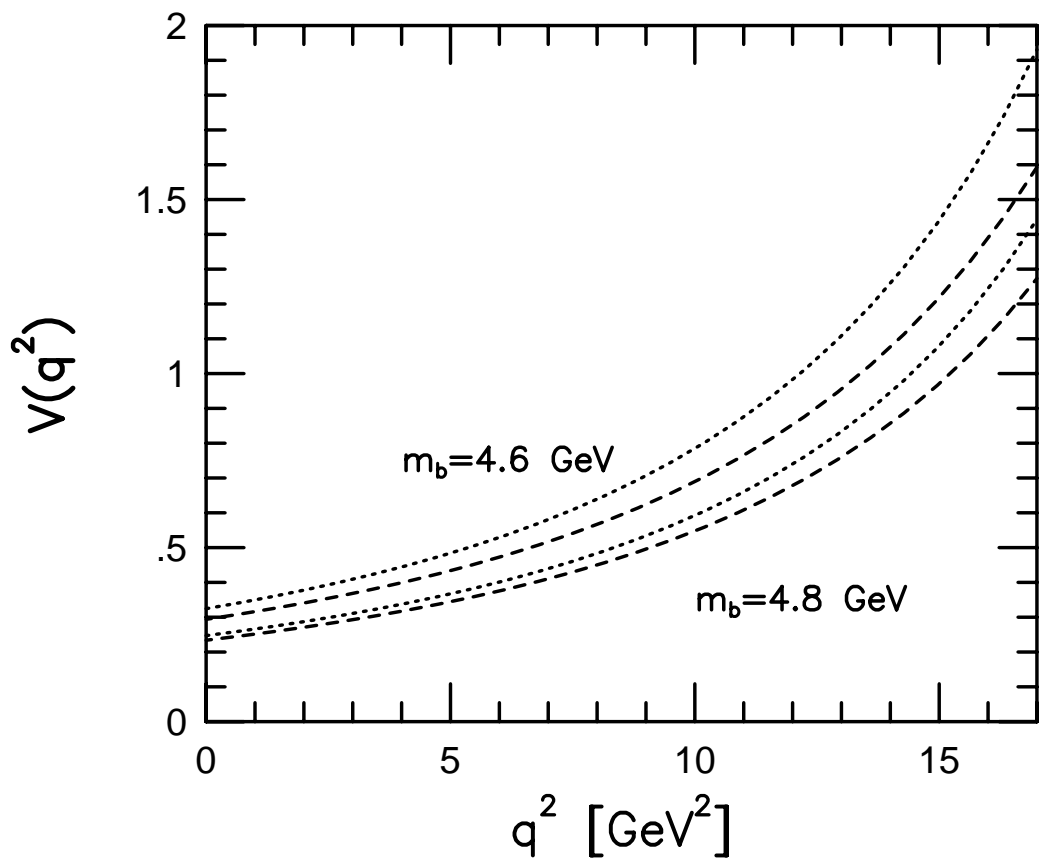
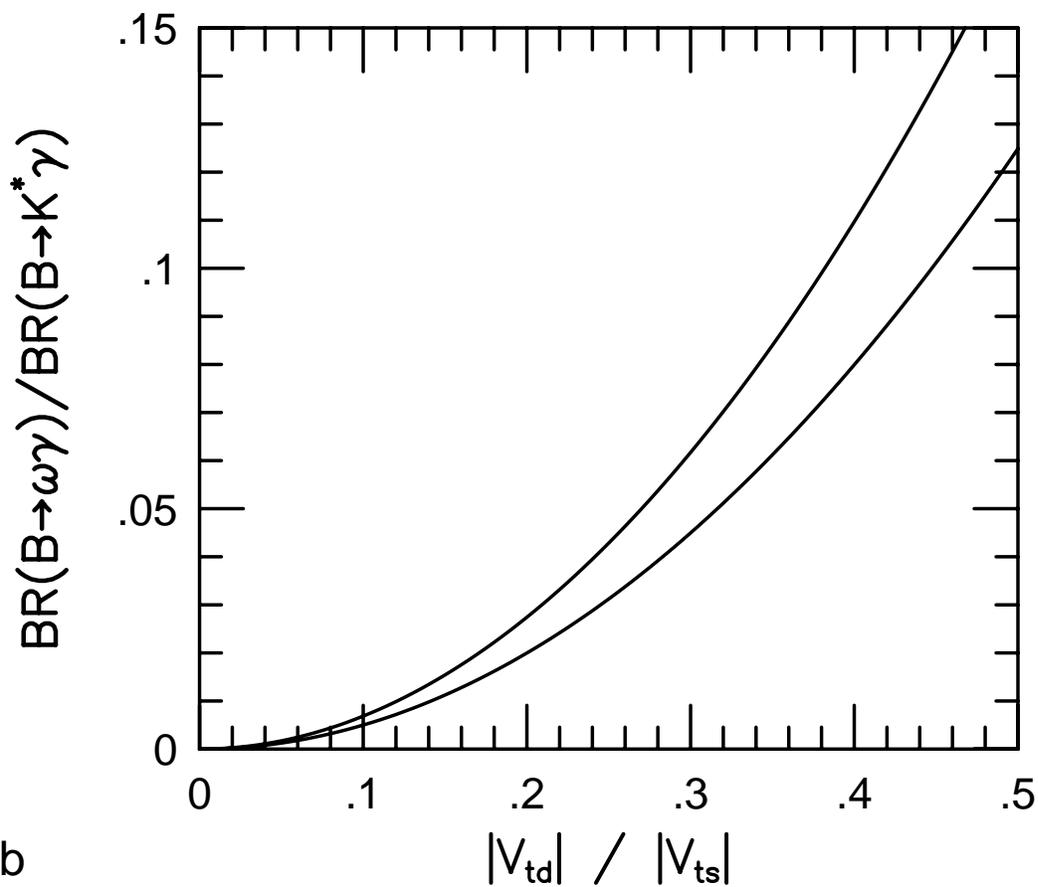
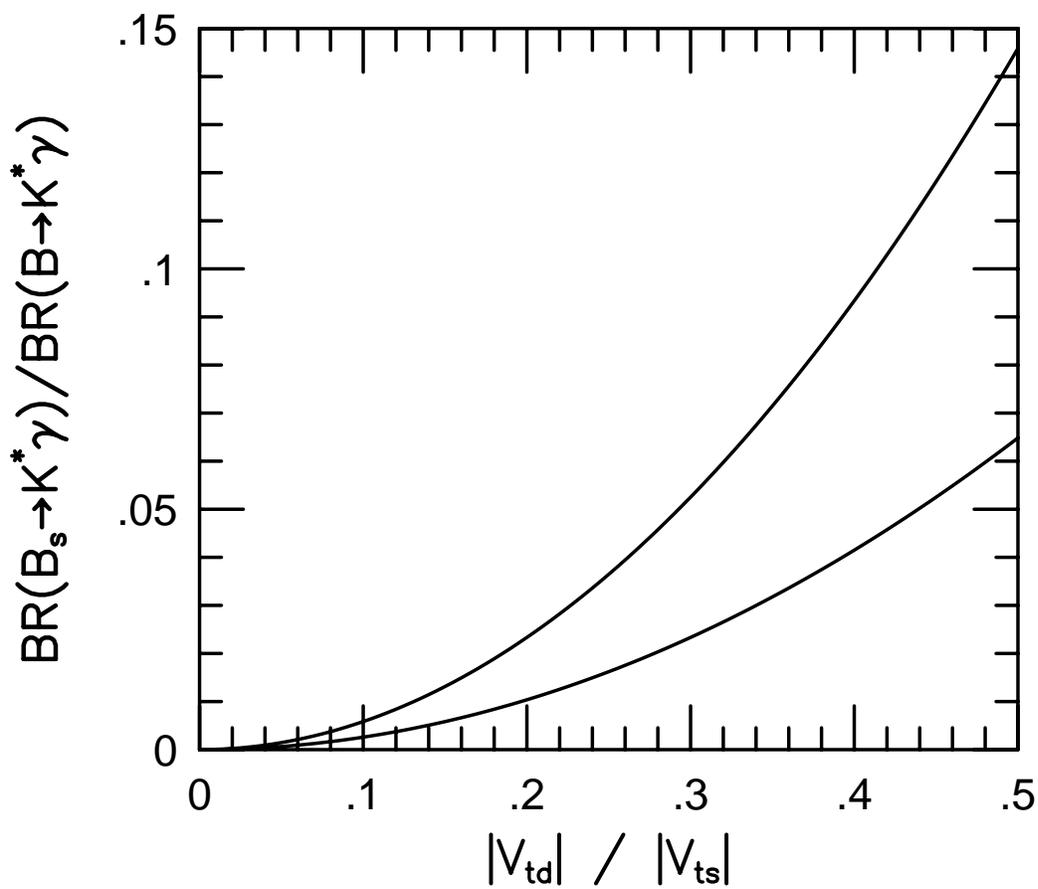
\end{document}